\pdfoutput=1
\documentclass[useAMS,usenatbib]{mn2e}
\usepackage{natbib}
\usepackage{graphicx}
\usepackage{amssymb}
\usepackage{amsmath}
\usepackage{calc}
\usepackage{color}
\usepackage[colorlinks=true,linkcolor=magenta,citecolor=blue]{hyperref}
\usepackage{ifpdf}

\title[Absorption processes in white dwarf atmospheres]
{The absorption non-symmetric ion-atom processes in the helium-rich white dwarf atmospheres}

\author[Ignjatovi{\' c} et al.]{Lj. M. Ignjatovi{\' c}$^{1,2}$ \thanks{E-mail: ljuba@ipb.ac.rs}, A. A. Mihajlov$^{1,2}$, V. A. Sre{\'c}kovi{\' c}$^{1,2}$
\newauthor and M. S. Dimitrijevi{\'c}$^{2,3,4,5}$\\
$^{1}$University of Belgrade,Institute of Physics, P. O. Box 57, 11001 Belgrade,
Serbia\\
$^{2}$Isaac Newton Institute of Chile, Yugoslavia Branch, Volgina 7, 11060 Belgrade, Serbia\\
$^{3}$Astronomical Observatory, Volgina 7, 11160 Belgrade 74,
Serbia\\
$^{4}$IHIS-Technoexperts, Be\v zanijska 23, 11080 Zemun, Serbia\\
$^{5}$Observatoire de Paris, 92195 Meudon Cedex, France\\
}

\begin{document}

\date{}

\pagerange{\pageref{firstpage}--\pageref{lastpage}} \pubyear{2008}

\maketitle

\label{firstpage}

\begin{abstract}

In this work the processes of absorption charge-exchange and photo-association
in He+H$^{+}$ collisions together with the process of ion HeH$^{+}$
photo-dissociation are considered as factors of influence on the opacity of the
atmospheres of helium-rich white dwarfs in the far UV and EUV region. It is shown
that they should be taken into account even in the cases of the atmospheres of
white dwarfs with H:He =$10^{-5}$. Than, it is established that in the cases of
white dwarfs with H:He $\gtrsim 10^{-4}$, particulary when H:He $\approx 10^{-3}$,
these processes have to be included \emph{ab initio} in the corresponding models of
their atmospheres, since in the far UV and EUV region they become dominant with respect to
the known symmetric ion-atom absorption processes.
\end{abstract}


\section{Introduction}

It has been shown recently in \citet{mih13}, that in order to
consider the contribution of the absorbtion processes
connected with binary ion-atom systems to the opacity of
the solar photosphere it is not enough to take into account only
the processes of absorbtion charge exchange in (H +
H$^{+}$)-collisions and the molecular ion H$_{2}^{+}$
photo-dissociation. These processes were studied in
\citet{mih86} and \citet{mih93a, mih94b, mih07a}, and are
already included in one of the solar photosphere models
\citep{fon09}. It has been established that in the very important
far UV and EUV spectral regions they have to be considered
together with the processes of the absorbtion charge exchange
and photo-association in non-symmetric (H + X$^{+}$)-collisions
and molecular ion H$X^{+}$ photo-dissociation, where $X$ is
one of metal atoms. Namely, it has been proved that only in
such case the total efficiency of the ion-atom absorbtion
processes in the mentioned spectral regions approaches the
efficiency of the relevant concurrent processes in the whole
solar photosphere.

\begin{figure}
\begin{center}
\includegraphics[width=\columnwidth,
height=0.75\columnwidth]{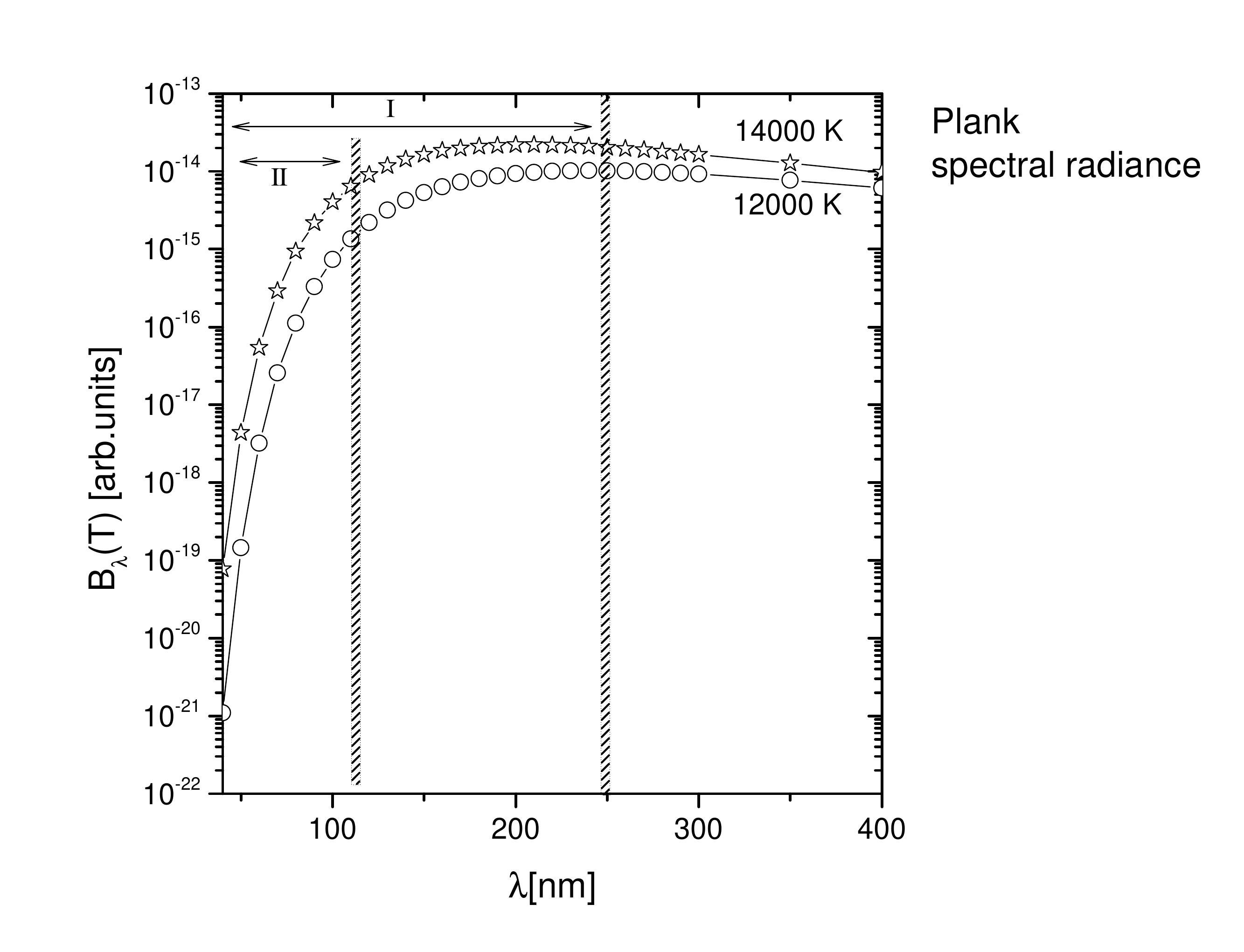} \caption{Plank curve for $T_{eff.}=12000$ K and
$T_{eff.}=14000$ K. "I" and "II" denote the regions $50.44$ nm $\le \lambda \lesssim
250$ nm and $50.44$ nm $\le \lambda \lesssim 120$ nm respectively. }
\label{fig:plank}
\end{center}
\end{figure}

These results suggest that it is useful to consider again the situation of
ion-atom absorbtion processes in the atmospheres of helium-rich white dwarfs.
Let us remind that in the previous papers \citep{mih92b, mih94a, sta94, mih95,
ign09}, dedicated to certain DB white dwarf atmospheres, the processes
of molecular ion He$_{2}^{+}$ photo-dissociation were studied:
\begin{equation}
\label{eq:sim1} \varepsilon_{\lambda} + He_{2}^{+}
\longrightarrow He + He^{+},
\end{equation}
and absorbtion charge exchange in (He + He$^{+}$)-collisions:
\begin{equation}
\label{eq:sim2} \varepsilon_{\lambda}
+ He^{+} + He \longrightarrow He + He^{+}
\end{equation}
where $\varepsilon_{\lambda}$ is the energy of a photon with wavelength
$\lambda$, He = He$(1s^{2})$, He$^{+}$ = He$^{+}(1s)$ and He$_{2}^{+}$ =
He$_{2}^{+}(1^{2}\Sigma_{u}^{+})$. The significance of these symmetric ion-atom
absorption processes for the atmospheres of the considered DB white dwarfs was
established in \citet{mih94a, mih95} and \citet{ign09} by a direct comparison of their
efficiencies with the main concurrent process of inverse "bremssthrallung" in
(free electron + He)-collisions, i.e.
\begin{equation}
\label{eq:e-He}
\varepsilon_{\lambda} + e + He \longrightarrow e' + He,
\end{equation}
where $e$ and $e'$ denote a free electron in the initial and final energetic
states respectively. For that purpose the data from the corresponding DB white-dwarf
atmosphere models \citep{koe80} have been used. It was established that the
processes (\ref{eq:sim1}) and (\ref{eq:sim2}) significantly influence the opacity
of the considered DB white dwarf atmospheres, with an effective temperature
$T_{eff} \ge 12000$ K, which fully justifies their inclusion in one of the models of
such atmospheres \citep{ber95}. However, the same comparison demonstrated also
that the dominant role in those atmospheres generally still belongs to the
concurrent absorbtion process (\ref{eq:e-He}), while the processes (\ref{eq:sim1})
and (\ref{eq:sim2}) can be treated as dominant (with respect to this concurrent
process) only in some layers of that atmospheres, and only within the part
50 nm$ < \lambda < $ 250 nm of the far UV and EUV region. In Fig. \ref{fig:plank},
where Plank's curves for $T_{eff}=12000$ K and $14000$ K are shown, this part is
denoted by $"I"$. Its boundary (from the short-wavelength side) is determined by
the value of wavelength $\lambda_{He} \approx 50.44$ nm,
which corresponds to the threshold of the atom He photo-ionization. Hence it follows
that in the case of helium-rich white dwarf atmospheres it would certainly be
useful to include into consideration some new ion-atom absorbtion processes,
which is principally allowed in accordance with the composition of such atmospheres
\citep{bue70}.

Let us note in this context that in the case of white dwarf atmospheres with
dominant helium component, among all possible symmetric ion-atom absorbtion
processes which are allowed by their composition \citep{bue70}, only the processes
(\ref{eq:sim1}) and (\ref{eq:sim2}) have to be taken into account. This means that
in this case we can find new relevant absorbtion processes only among the processes
connected with non-symmetric ion-atom systems, particulary such systems which
could provide efficiency in the same part $"I"$ of the far UV and EUV region.
Here we will examine the significance of non-symmetric ion-atom absorption
processes with participation of the hydrogen component. We mean the processes of
molecular ion HeH$^{+}$ photo-dissociation
\begin{equation}
\label{eq:nonsim1} \varepsilon_{\lambda}
+ HeH^{+} \longrightarrow He^{+} + H
\end{equation}
and the processes of absorption charge exchange and photo-association
in the (He + H$^{+}$)-collisions, namely
\begin{equation}
\label{eq:nonsim2} \varepsilon_{\lambda}
+ He + H^{+} \longrightarrow He^{+} + H,
\end{equation}
\begin{equation}
\label{eq:nonsim3} \varepsilon_{\lambda}
+ He + H^{+} \longrightarrow (HeH^{+})^{*} ,
\end{equation}
where H = H$(1s)$, He= He$(1s^{2})$, He$^{+}$ = He$^{+}(1s)$, and HeH$^{+}$ and
(HeH$^{+}$)$^{*}$ denote the molecular ion in the ground and the first excited electronic
states which are adiabatically correlated with the states of the systems He +
H$^{+}$ and He$^{+}$ + H respectively at the infinite internuclear distance. Already
in \citet{mih13} it was noted that these processes, whose significance was
practically neglected for the solar photosphere, could be rather important in the
case of helium-rich white dwarf atmospheres. This assumption was worthy of
attention, particulary due to the fact that characteristics of the
considered non-symmetric molecular ions (see Fig.\ref{fig:mol.ion}) provide
manifestations of the processes (\ref{eq:nonsim1}) - (\ref{eq:nonsim3}) just in
the part $"I"$ of the far UV and EUV region.

In connection with this fact, let us remind that the part $"I"$ is rather important
for such values of $T_{eff}$. Namely, let $\lambda_{max}$ be the positions of the
maxima of the spectral intensities characterizing the electromagnetic (EM) emission
of the considered atmospheres, which are determined from the well-known Wien's law:
$\lambda_{max} \cdot T_{eff} = 2.898 \cdot 10^{6}$ nm $\cdot$ K. We then have
it that for the considered DB white dwarfs $\lambda_{max} < 250$ nm, so that the
mentioned maxima lie just within the region $"I"$.

It was just because of the above mentioned that this investigation was undertaken.
The main aim was to study when the non-symmetric absorbtion processes (\ref{eq:nonsim1}) -
(\ref{eq:nonsim3}) can significantly influence the opacity of helium-rich
white dwarf atmospheres in the part $"I"$ of the UV and VUV spectral region, and
to show that the processes (\ref{eq:nonsim1}) - (\ref{eq:nonsim3}) deserve to
be included \emph{ab initio} in the corresponding white dwarf atmosphere models.
Therefore the relevant spectral characteristics of the processes (\ref{eq:nonsim1})
- (\ref{eq:nonsim3}) are determined here for the atmospheres of
different helium-rich white dwarfs with $T_{eff}=12000$ K and $14000$ K,
log $g=8$ and $7$, and for the values of the ratio H:He from $10^{-5}$ to $10^{-3}$.
The necessary expressions for these spectral characteristics are given in
Section 2. Then, with their help, in the Section 3 the values are calculated of the
parameters characterizing the relative efficiency of the non-symmetric processes
(\ref{eq:nonsim1}) - (\ref{eq:nonsim3}) with respect to the efficiency of all ion-atom
processes, as well as with respect to the electron-atom process (\ref{eq:e-He}), which
comprise the main direct results of this work.

Let us note that already in \citep{ign09}, beside the electron-atom processes
(\ref{eq:e-He}) the process of photoionization of
hydrogen atoms, was also considered, namely
\begin{equation}
\label{eq:phi}
\varepsilon_{\lambda} + H \longrightarrow e + H^{+},
\end{equation}
where H=H$(1s)$. It was treated as a concurrent process, potentially
necessary in the region $\lambda < \lambda_{H}$ where $\lambda_{H}\approx 911$
{\AA} corresponds to the threshold of atom H photoionization.
However, in \citep{ber13} it was brought to attention that the
importance of this absorption channel was significantly underestimated.
That is why in this work the process (\ref{eq:phi}) was again included
into the consideration and carefully examined.

One can see that in this work we take into account only such non-symmetric
ion-atom absorption processes where, apart from the dominant helium component of the
considered atmospheres, only the hydrogen component participates, although they contain also a
lot of metal components \citep{bue70}. This is due to the fact that the existing
atmosphere models do not provide the necessary data (about the relevant metals'
abundances) which would be needed for the present calculations. However, we consider that
a demonstration of the fact that for the considered atmospheres the
processes (\ref{eq:nonsim1}) - (\ref{eq:nonsim3}), where one of
their "minor" components participates, are rather significant, is
a sufficient reason for treating the non-symmetric
ion-atom absorption processes in general as potentially
significant for those atmospheres.

\section{The theoretical remarks: the relevant spectral characteristics}

\subsection{The non-symmetric ion-atom processes}
As the relevant characteristics of the processes
(\ref{eq:nonsim1}), (\ref{eq:nonsim2}) and (\ref{eq:nonsim3})
we will use the corresponding spectral absorption coefficients.
They are defined as functions of log $\tau$, where $\tau$
is Rosseland optical depth of the part of the examined
atmosphere above the considered layer for the wavelength
$\lambda$. They are denoted here as
$\kappa^{(bf)}_{nsim}(\lambda;\log \tau)$,
$\kappa^{(ff)}_{nsim}$ $(\lambda;\log \tau)$ and
$\kappa^{(fb)}_{nsim}(\lambda;\log \tau)$, in accordance with
the fact that the mentioned processes can be treated as
bound-free, free-free and free-bound respectively. These
coefficients are determined here within the corresponding atmosphere models,
by means of the local temperature and the densities of He atoms and H$^{+}$ ions,
and used in a similar form, namely
\begin{equation}
\label{eq:kapabf}
\kappa^{(bf,ff,fb)}_{nsim}(\lambda;\log \tau) = K^{(bf,ff,fb)}_{nsim}(\lambda;T)
N_{He}N_{H^{+}},
\end{equation}
where $T\equiv T(\log \tau)$, $N_{He}\equiv N_{He}(\log \tau)$ and $N_{H^{+}}\equiv
N_{H^{+}}(\log \tau)$. Of course, it is understood that the photo-dissociation rate
coefficient $K_{nsim}^{(bf)}(\lambda;T)$ is given by the known relations
\begin{equation}
\label{eq:Kbfphd}
K^{(bf)}_{nsim;X}(\lambda;T)=\sigma^{(phd)}_{HeH^{+}}(\lambda,T)
\cdot \chi^{-1}(T;HeH^{+}),
\end{equation}
\begin{equation}
\label{eq:Kiaa}
\chi(T;HeH^{+})=\left [\frac{N_{He}\cdot N_{H^{+}}}{N_{HeH^{+}}} \right],
\end{equation}
where $\sigma^{(phd)}_{HeH^{+}}(\lambda,T)$ is the mean thermal cross-section for
the molecular ion HeH$^{+}$ photo-dissociation, $N_{HeH^{+}}$ denotes the local
density of these molecular ions, and $\chi(T;HeH^{+})$ is determined under
condition of local thermodynamical equilibrium (LTE) with given $T$, $N_{He}$
and $N_{H^{+}}$. Let us note that these expressions contain no correction
factors which take into account the influence of stimulated emission,
since within the actual range of $\varepsilon_{\lambda}/kT$ ratio for the considered
cases the changes due to these factors will be of the order of magnitude of $10^{-3}\%$.
\begin{figure}
\begin{center}
\includegraphics[width=\columnwidth,
height=0.75\columnwidth]{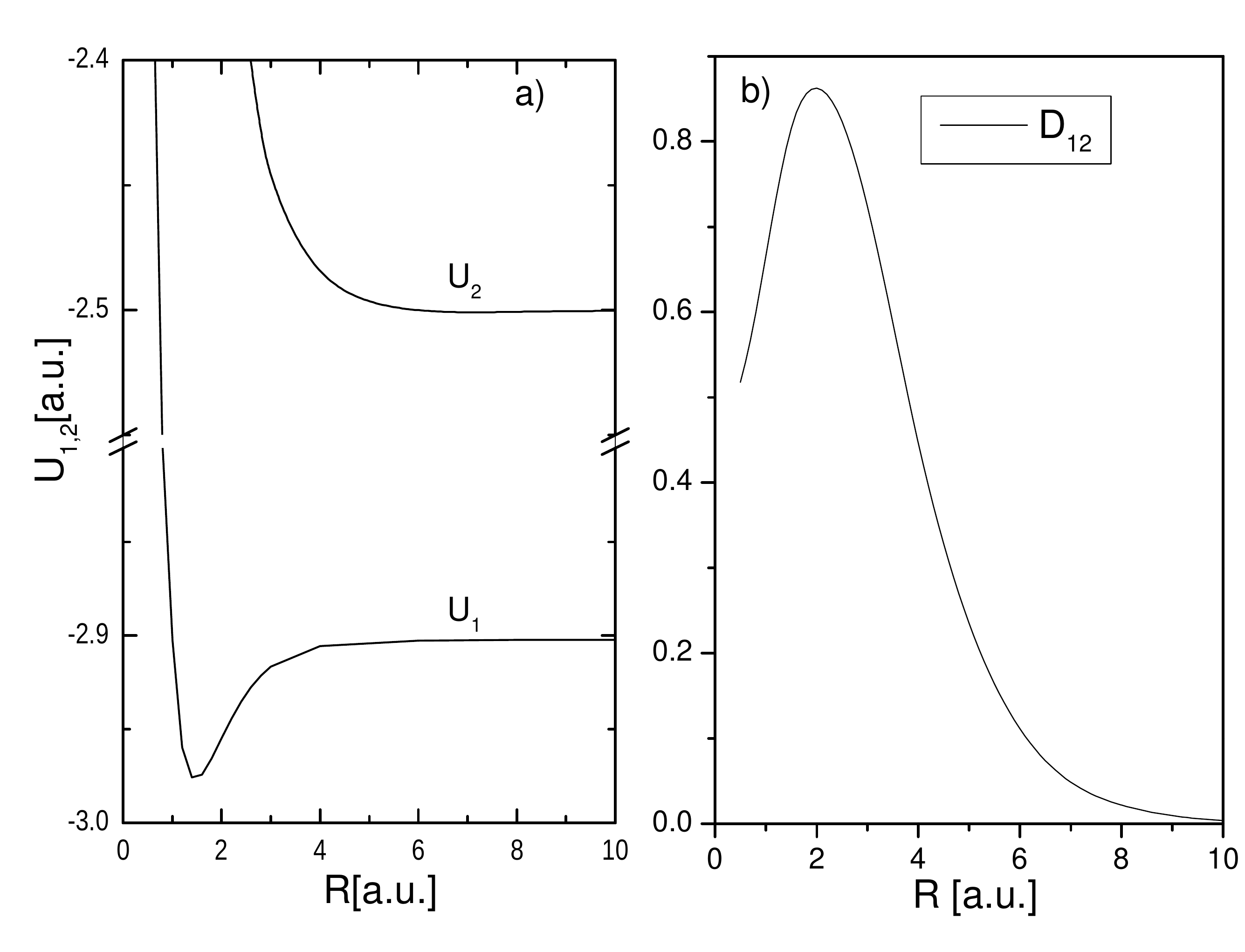} \caption{The potential curves $U_{1}(R)$
and $U_{2}(R)$ for the ground and the first excited electronic states of the
molecular ion HeH$^+$ and the corresponding dipole matrix element $D_{12}(R)$, where
$R$ is the internuclear distance.}
\label{fig:mol.ion}
\end{center}
\end{figure}

Finally, the total efficiency of the non-symmetric processes
(\ref{eq:nonsim1}), (\ref{eq:nonsim2})  and (\ref{eq:nonsim3}) is characterized here
by the spectral absorption coefficient $\kappa_{nsim}(\lambda;\log \tau)$ given by
the relations
\begin{equation}
\label{eq:kapansim}
\kappa_{nsim}(\lambda;\log \tau)= K_{nsim}(\lambda;T)N_{He} N_{H^{+}},
\end{equation}
\begin{equation}
\label{eq:Knsim}
K_{nsim}(\lambda;T)=K^{(bf)}_{nsim}(\lambda;T)+K^{(ff)}_{nsim}(\lambda;T)
+ K^{(fb)}_{nsim}(\lambda;T),
\end{equation}
where the rate coefficients $K^{(bf,ff,fb)}_{nsim}(\lambda,T)$ are determined by
means of the necessary characteristics of the considered molecular ion, in a way which
was described in detail in \citet{mih13}.

The mentioned characteristics, i.e. the adiabatic potential curves of the
molecular ion in the ground state (HeH$^{+}$) and the first excited electronic state
((HeH$^{+}$)$^{*}$), as well as the corresponding dipole matrix element (which were
not shown in \citet{mih13}) are presented here in Fig.\ref{fig:mol.ion} as
functions of the internuclear distance $R$. This figure also shows in a schematic way
the bound-free (bf), free-free (ff) and free-bound (fb) transitions between
the energy states of the considered ion-atom system which correspond to the
processes (\ref{eq:nonsim1}), (\ref{eq:nonsim2}) and (\ref{eq:nonsim3}). The
potential curves are denoted in Fig. \ref{fig:mol.ion} by $U_{1}(R)$ and
$U_{2}(R)$, and dipole matrix element - by $D_{12}(R)$. Their values as
functions of $R$ are determined here by fitting the corresponding data from
\citet{gre74a} and \citet{gre74b}.

Let us note that these data have a shortcoming: they are not sufficiently complete
and do not guarantee that different ways of fitting give close values of
$U_{1}(R)$, $U_{2}(R)$ and $D_{12}(R)$. It is possible that this shortcoming
causes the observed differences between our values of the partial cross-section for
photo-dissociation of the ion HeH$^{+}$ from its ground ro-vibrational state and
the values presented in \citet{dum09}, where they were calculated by means of the
same data. However, here we used just the data from \citet{gre74a} and
\citet{gre74b} since as yet only these papers give at least some data about both the
necessary potential curves (of the ion HeH$^{+}$) and the corresponding matrix element.
Also, we keep in mind that development of numerical procedures which would
be suitable for improvement of the data presented in \citet{gre74a} and
\citet{gre74b} far exceeds the aim of this work. Apart from that, we consider that the
deviations of the potential curves $U_{1}(R)$ and $U_{2}(R)$ from the hypothetical
exact ones do not cause any large errors in the obtained results and consequently
cannot strongly influence the final conclusions.

Since all processes (\ref{eq:nonsim1}) - (\ref{eq:nonsim3}) are connected with the
transition between the ground and the first excited electronic states of the strongly
non-symmetric ion-atom system (HeH$^{+}$ or He + H$^{+}$) we have it that the range of
values of the splitting term ($U_{12}(R) \equiv U_{2}(R) - U_{1}(R)$) well
characterizes the range of photon energies which is relevant for those
processes. It is just due to this fact that we could state above that the processes
(\ref{eq:nonsim1}) - (\ref{eq:nonsim3}) have to be manifested in the part $"I"$
of the far UV and EUV spectral region (see Fig. \ref{fig:plank}).

\begin{figure}
\begin{center}
\includegraphics[width=\columnwidth,
height=0.75\columnwidth]{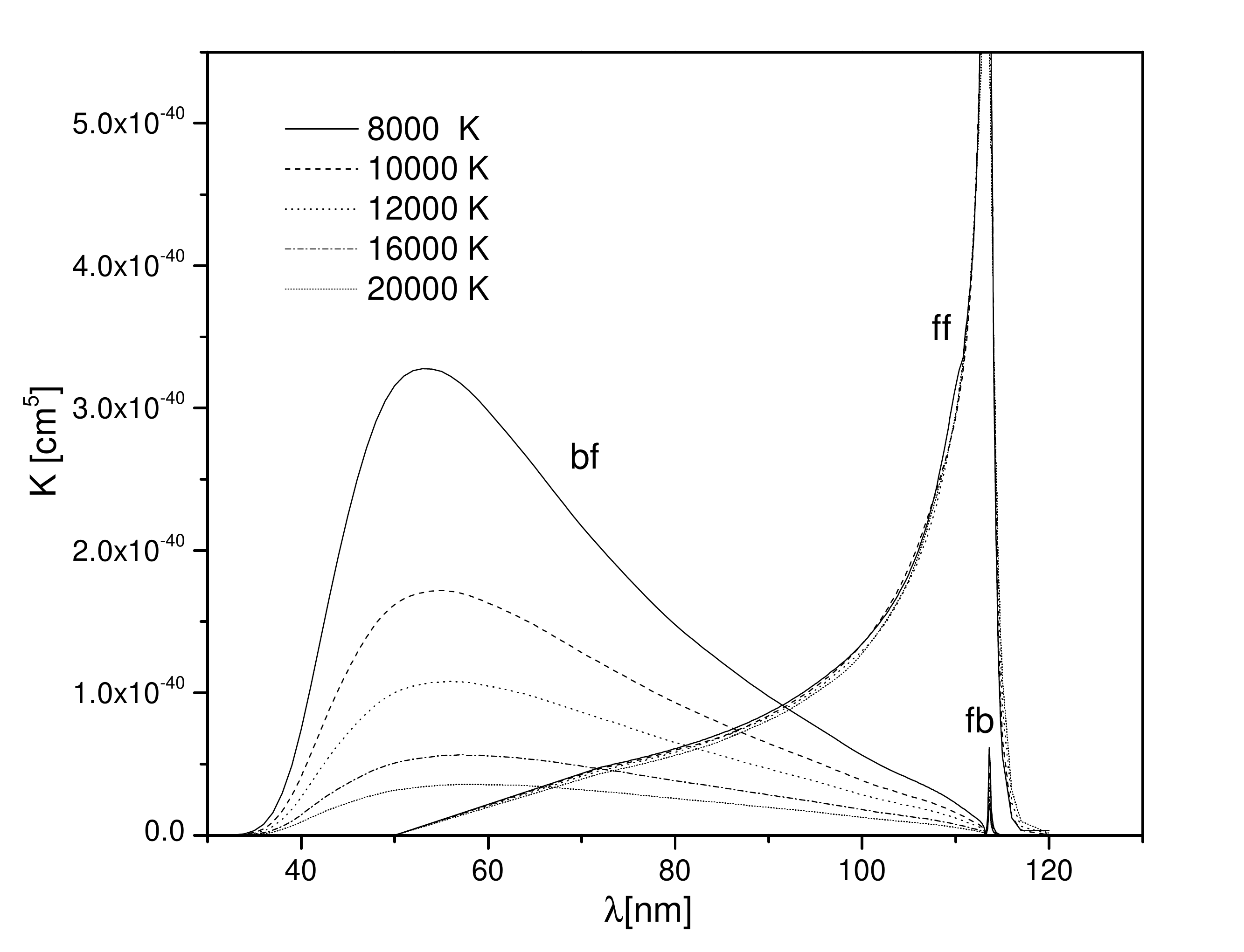} \caption{The behavior of the
bound-free (bf), free-free (ff) and free-bound (fb) spectral rate coefficients
$K^{(bf,ff,fb)}_{HeH^{+}}(\lambda;T)$ for the molecular ion HeH$^{+}$.}
\label{fig:Kbffffb}
\end{center}
\end{figure}

Although the behavior of the rate coefficients $K^{(bf,ff,fb)}_{nsim}(\lambda,T)$
was already discussed in \citet{mih13}, it is also illustrated here by
Fig. \ref{fig:Kbffffb} since the range of
temperatures characterizing the solar photosphere is not relevant in our case.
These figures give a possibility to estimate that the processes (\ref{eq:nonsim1}) -
(\ref{eq:nonsim3}) can be significant in the spectral region denoted in Fig.
\ref{fig:plank} by $"II"$.

\subsection{The symmetric ion-atom, electron-atom and hydrogen photo-ionization processes}
For the sake of the following considerations we have to introduce the spectral coefficients
$\kappa_{sim}(\lambda;\log \tau)$, $\kappa_{e-He}(\lambda;\log \tau)$ and
$\kappa_{phi}(\lambda;\log \tau)$ which characterize the efficiencies of the
symmetric ion-atom absorption processes (\ref{eq:sim1}) and (\ref{eq:sim2}) together,
the electron-atom processes (\ref{eq:e-He}) and hydrogen photoionization process (\ref{eq:phi})
respectively. As in \citet{ign09} we can take these coefficients in the known form:
\begin{equation}
\label{eq:kapasim}
\kappa_{sim}(\lambda;\log \tau) = K_{sim}(\lambda,T) \cdot N_{He} \cdot N_{He^{+}},
\end{equation}
\begin{equation}
\label{eq:kapae-He}
\kappa_{e-He}(\lambda;\log \tau) = K_{e-He}(\lambda,T) \cdot N_{He} \cdot N_{e},
\end{equation}
\begin{equation}
\label{eq:kapa-phi}
\kappa_{phi}(\lambda;\log \tau) = \sigma_{phi}(\lambda;H) \cdot N_{H},
\end{equation}
where $N_{He^{+}}$, $N_{e}$ and $N_{H}$ are the local densities of ions He$^{+}$,
free electrons and atoms H, $K_{sim}(\lambda,T)$ and $K_{e-He}(\lambda,T)$ - adequately
defined spectral rate coefficients, and $\sigma_{phi}(\lambda;H)$ - spectral
cross-section for atom H photoionization. The absorption coefficient $K_{sim}(\lambda,T)$
is determined in the way which is described in detail in \citet{ign09}, and $K_{e-He}(\lambda,T)$
 - by means of the data from \citet{som65}, and $\sigma_{phi}(\lambda;H)$ is taken from \citep{bet57}.

Here we take into account the fact that all the values of the rate coefficients
$K_{sim}(\lambda,T)$ and $K_{e-He}(\lambda,T)$ which are needed for our calculations
have already been determined in \citet{ign09}, and that the photo-ionization cross-section
$\sigma_{phi}(\lambda;H)$ is given by a known analytical expression. Therefore
in the further text we will simply treat these characteristics and, consequently, the
spectral absorption coefficients $\kappa_{sim}(\lambda,\log \tau)$,
$\kappa_{e-He}(\lambda,\log \tau)$ and $\kappa_{phi}(\lambda;\log \tau)$ as known quantities.

\section{Results and Discussion}

\subsection{DB white dwarfs}
As it is well known, the above defined spectral absorption
coefficients depend on the wavelength, local temperature, and local particle densities,
based on the corresponding models of
helium-rich white dwarf atmospheres characterized by certain values of $T_{eff}$,
$\log g$ and the ratio of the hydrogen and helium species (H:He). Here we will start from
DB white dwarf atmospheres with $T_{eff}=12000$ K and $14000$ K, $\log g=8$ and
$7$ and H:He $= 10^{-5}$. For their description we will use the equilibrium models
which are presented in \citet{koe80}. As in \citet{mih94a, mih95}, and
\citet{ign09}, it is due to the fact that, although newer atmosphere
models for helium-rich white dwarfs now exist (see e.g. the review article of
\citet{koe10}), only the models from \citet{koe80} contain in a tabular form all
the relevant data which are needed for our calculations.

\begin{figure}
\begin{center}
\includegraphics[width=\columnwidth,
height=0.75\columnwidth]{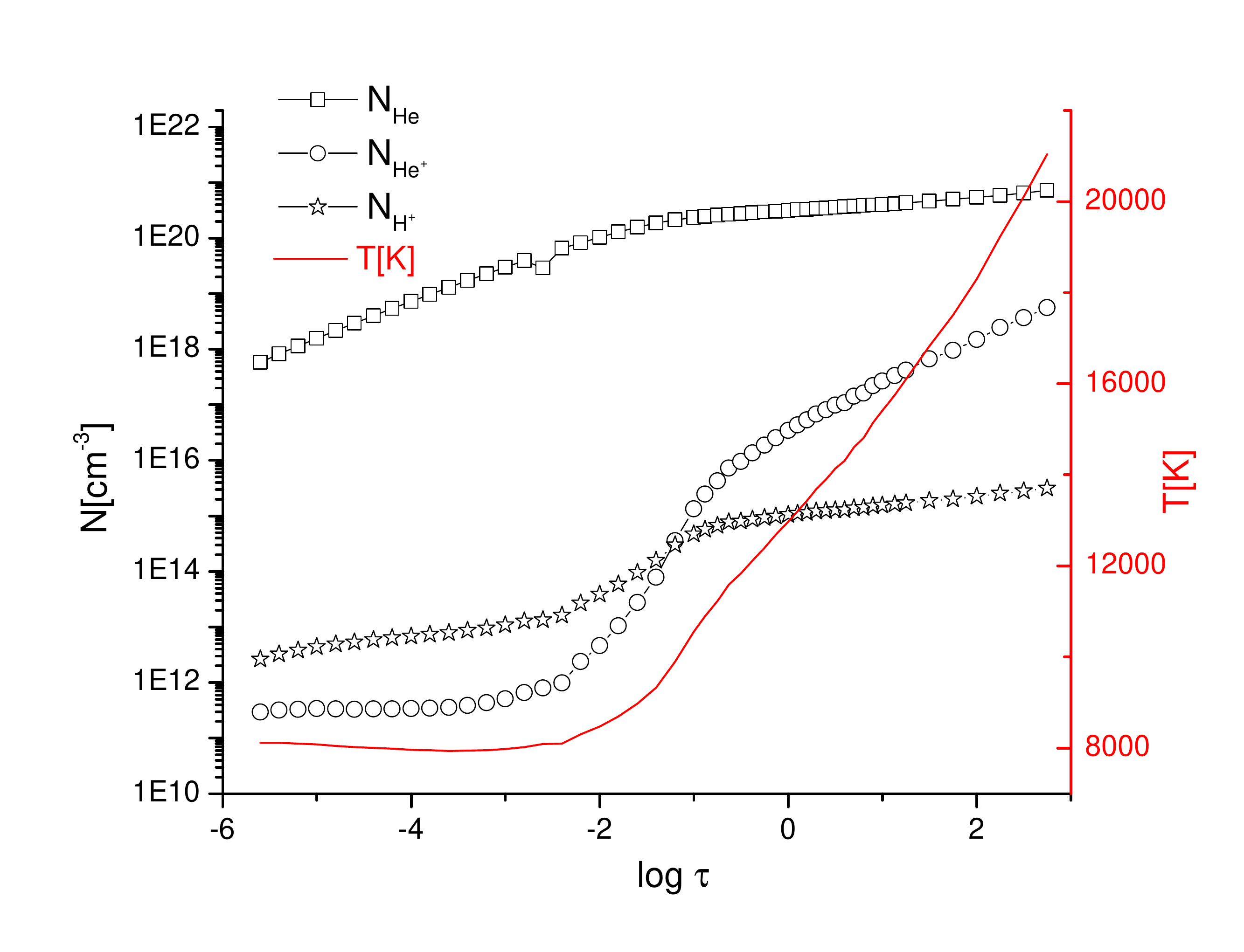} \caption{The local densities $N_{He}$,
$N_{He^+}$ and $N_{H^+}$ and the temperature $T$ as functions of $\log \tau$, where
$\tau$ is Rosseland  optical depth, according to the model of the DB white dwarf
atmosphere from \citep{koe80} for: $T_{eff}$=12000 K, $\log g=8$ and
H:He=$10^{-5}$.}
\label{fig:abundDB}
\end{center}
\end{figure}

The behavior of the densities of free electrons and ions He$^{+}$ and H$^{+}$
in the atmosphere of a DB white dwarf with $T_{eff}=12000$ K and log $g=8$ is
illustrated by Fig. \ref{fig:abundDB}, which shows that the processes
(\ref{eq:nonsim1}) - (\ref{eq:nonsim3}) could be of interest already in the case
H:He $= 10^{-5}$. Namely, this figure suggests that for $T_{eff} \lesssim 14000$ K,
the ion H$^{+}$ density is even larger than that of He$^{+}$ in significant parts of
DB white dwarfs' atmospheres ($\log \tau < -1$).

In accordance with the aim of this work we have to estimate first the relative
efficiency of the non-symmetric processes (\ref{eq:nonsim1}) - (\ref{eq:nonsim3}) with
respect to the total efficiency of all the above mentioned ion-atom absorption
processes, which is characterized by the spectral absorption coefficient
\begin{equation}
\label{eq:kapaia}
\kappa_{ia}(\lambda;\log \tau)= \kappa_{nsim}(\lambda;\log \tau) +
\kappa_{sim}(\lambda;\log \tau),
\end{equation}
For that purpose we use the quantity
\begin{equation}
\label{eq:Gnsim}
G^{(nsim)}_{ia}(\lambda;\log \tau) = \frac{\kappa_{nsim}(\lambda;\log \tau)}
{\kappa_{ia}(\lambda;\log \tau)}.
\end{equation}
One can see that the definition of this quantity guarantees the validity of
the relations $0 < G^{(nsim)}_{ia}(\lambda;\log \tau) < 1$ for any $\lambda$ and
$\log \tau$. It is important since other possible quantities, i.e.
$\kappa_{ia}(\lambda;\log \tau)/\kappa_{sim}(\lambda;\log \tau)$ and
$\kappa_{nsim}(\lambda;\log \tau)$/ $\kappa_{sim}(\lambda;\log \tau)$, could not
be practically presented in the corresponding figure.

\begin{figure}
\begin{center}
\includegraphics[width=\columnwidth,
height=0.75\columnwidth]{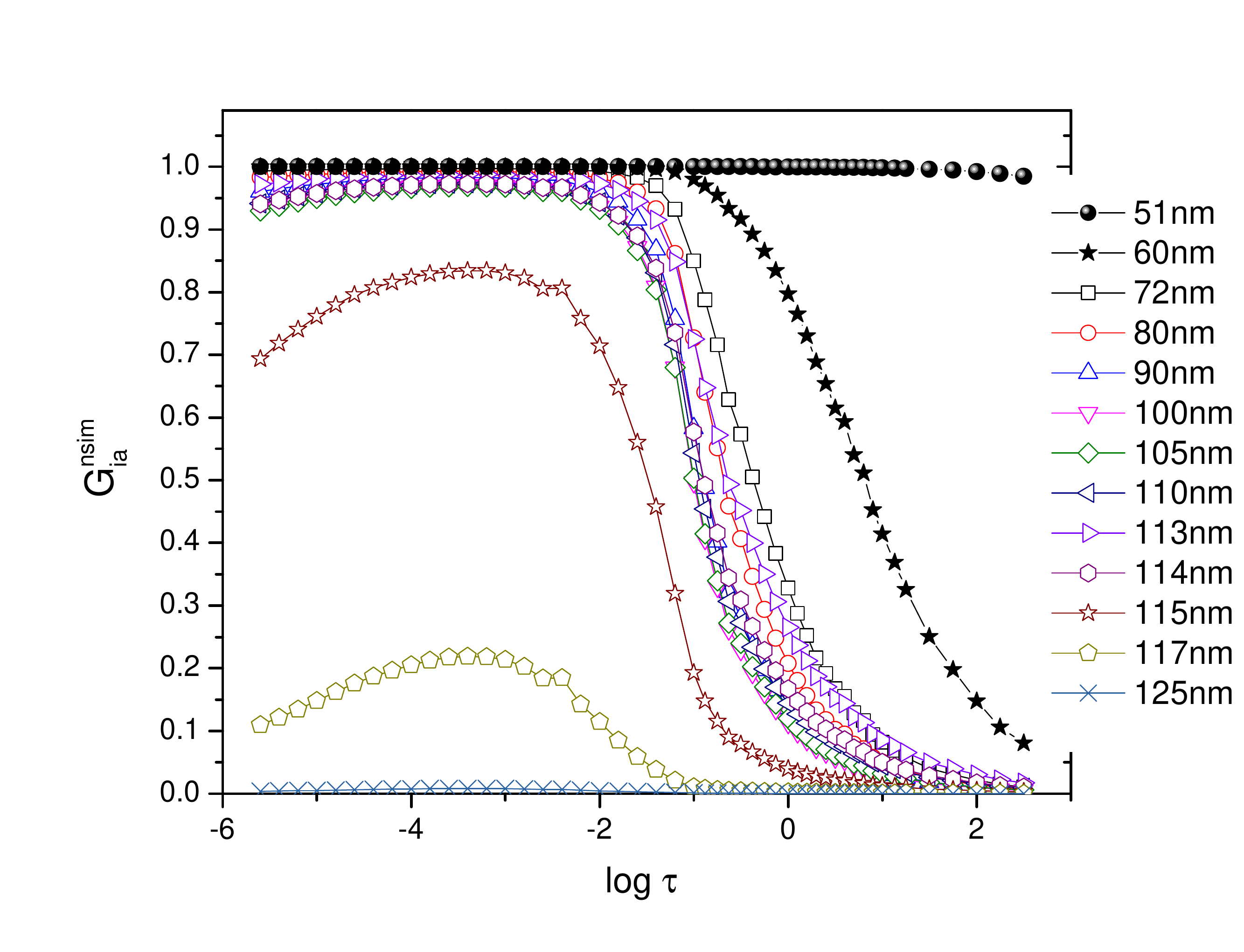} \caption{The behavior of the quantity
$G_{ia}^{nsim}=\kappa_{nsim}/\kappa_{ia}$, see Eq. (\ref{eq:Gnsim}), for DB white
dwarf atmosphere with $T_{eff}$=12000 K, $\log g=8$ and H/He=$10^{-5}$.}
\label{fig:G5}
\end{center}
\end{figure}
The behavior of the quantity $G^{(nsim)}_{ia}(\lambda;\log \tau)$ for a DB white dwarf
atmosphere with $T_{eff}=12000$ K, $\log g=8$ and H:He$= 10^{-5}$ is illustrated by
Fig. \ref{fig:G5}. This figure shows that the non-symmetric processes
(\ref{eq:nonsim1}) - (\ref{eq:nonsim3}) are dominant within a significant part of
the considered atmosphere ($-5.6\le$ $\log \tau \lesssim -0.75$), which corresponds to
the part in Fig. \ref{fig:abundDB} where $N_{H^{+}} > N_{He^{+}}$.

\begin{figure}
\begin{center}
\includegraphics[width=\columnwidth,
height=0.75\columnwidth]{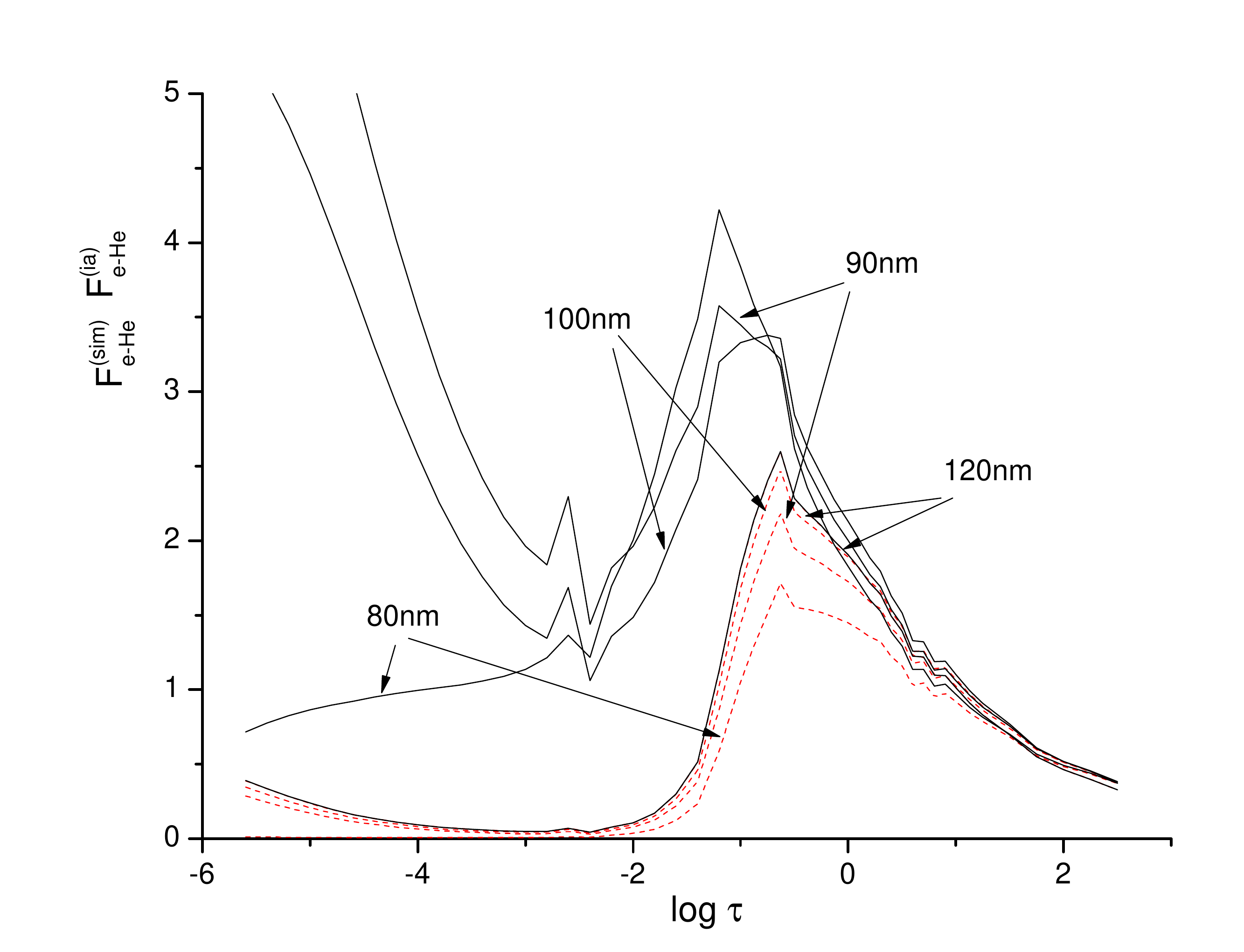} \caption{The behavior of the quantities
$F^{(sim)}_{e-He}(\lambda;\log \tau)=\kappa_{sim}/\kappa_{ea}$ (red line-dashed) and
$F^{(ia)}_{e-He}(\lambda;\log \tau)=\kappa_{ia}/\kappa_{ea}$ (black line-full), see Eq.
(\ref{eq:Fnsim}), for DB white dwarf atmosphere with $T_{eff}$=12000 K, $\log =8$
and H/He=$10^{-5}$.}
\label{fig:F5}
\end{center}
\end{figure}

In order to establish how the inclusion of the non-symmetric
processes (\ref{eq:nonsim1}) - (\ref{eq:nonsim3}) into the consideration influences the relative
efficiency of the ion-atom absorption processes with respect to the efficiency of the
concurent electron-atom process (\ref{eq:e-He}), we calculated the quantities
\begin{equation}
\label{eq:Fnsim}
\begin{split}
F^{(sim)}_{e-He}(\lambda;\log \tau) = \frac{\kappa_{sim}(\lambda;\log \tau)}
{\kappa_{e-He}(\lambda;\log \tau)}, \\
F^{(ia)}_{e-He}(\lambda;\log \tau) = \frac{\kappa_{ia}(\lambda;\log \tau)}
{\kappa_{e-He}(\lambda;\log \tau)}.
\end{split}
\end{equation}
Comparison of these two quantities gives a possibility to estimate the change of
the mentioned relative efficiency. The behavior of these quantities in the case of the
considered DB white dwarf atmosphere ($T_{eff}=12000$ K, log $g=8$, H:He$= 10^{-5}$) is shown
in Fig.\ref{fig:F5}. From this figure one can see that:\\
- the inclusion of the ion-atom non-symmetric absorption processes causes a very
significant increase in the relative efficiency of the ion-atom absorption processes
in just that region $\log \tau < 0.75$, i.e. where the symmetric processes can
be practically neglected.

Then, we established the fact that the behavior of the quantities
$G^{(nsim)}_{ia}(\lambda;\log \tau)$, $F^{(sim)}_{e-He}(\lambda;\log \tau)$ and
$F^{(ia)}_{e-He}(\lambda;\log \tau)$ is also similar in the cases of DB white dwarf
atmospheres with the same value of H:He, but with $T_{eff}=14000$ K and $\log g=8$,
and $T_{eff}=12000$ K and $\log g=7$. Based on the above mentioned, it
can be concluded that the non-symmetric processes (\ref{eq:nonsim1}) - (\ref{eq:nonsim3})
have a visible significance for the atmospheres of the considered DB white dwarfs with
H:He$= 10^{-5}$ and should be included in their models.

\begin{figure}
\begin{center}
\includegraphics[width=\columnwidth,
height=0.75\columnwidth]{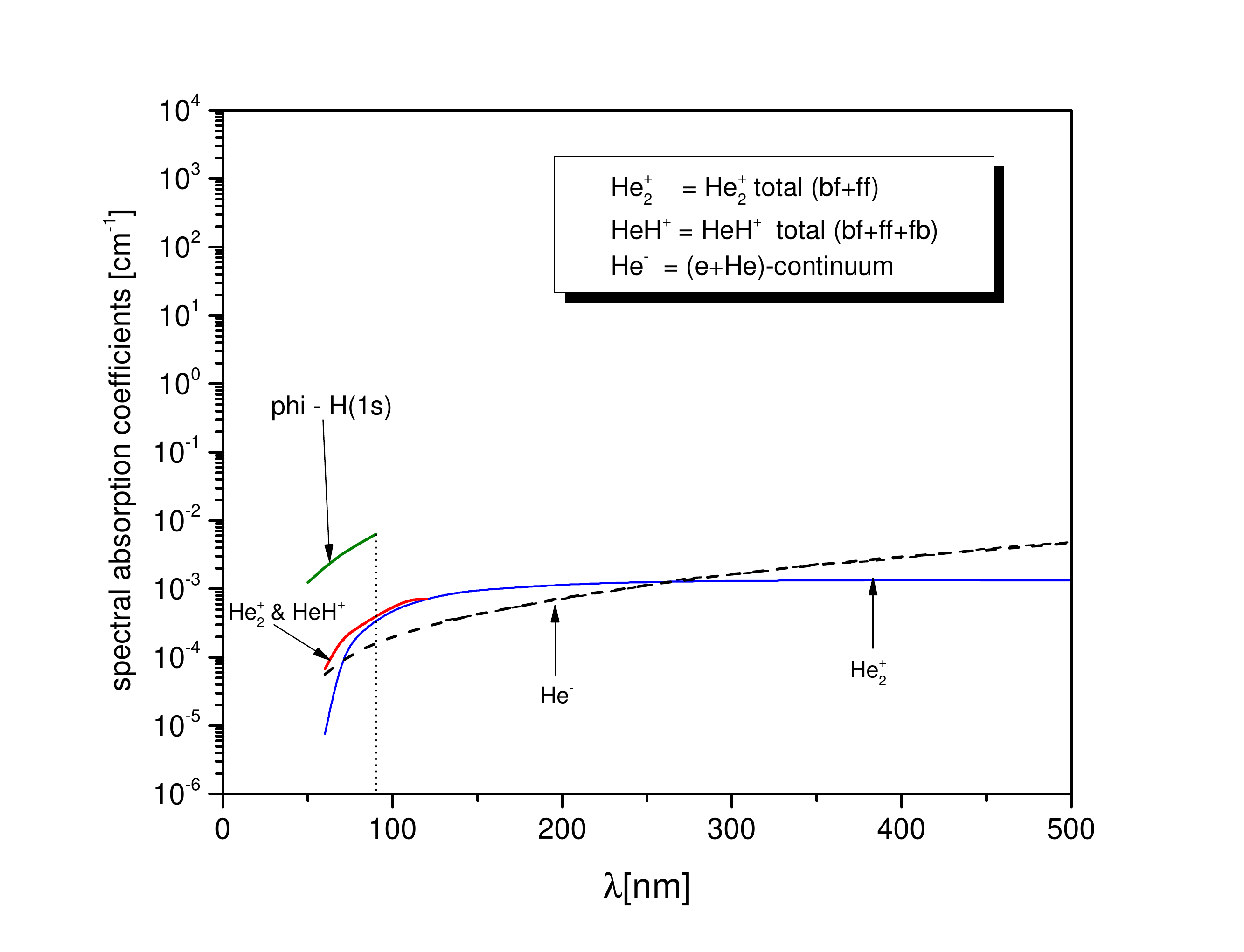} \caption{The plots of all considered absorbtion
processes for $\log \tau=0$ in the case of a DB white dwarf with $T_{eff}$=12000
K, log $g=8$ and H:He=$10^{-5}$.}
\label{fig:1}
\end{center}
\end{figure}

It is necessary to draw attention to the fact that this conclusion refers to a spectral
region $\lambda > \lambda_{H}$ where the hydrogen photo-ionization
process (\ref{eq:phi}) is impossible. In order to estimate the partial
efficiencies of the mentioned processes in the case of the considered
DB white dwarf in the whole region  $\lambda > \lambda_{He}$
the corresponding plots of all discussed absorption processes
for $\log \tau = 0$ are presented in Fig.\ref{fig:1} . One can see that in the region
$\lambda_{He} < \lambda < \lambda_{H}$ the process (\ref{eq:phi}) alone
gives the dominant contribution to the opacity of the considered atmosphere.
Here it was established that in the considered case this dominance
exists for any $\log \tau < 0$.

However, the main results of the research of DB white dwarf atmospheres is
the establishment of the fact that:\\
- the inclusion of the non-symmetric processes causes an increase of the
total efficiency of ion-atom absorption processes in the region $0.75 < \log
\tau < 2$, where the symmetric processes (\ref{eq:sim1}) and (\ref{eq:sim2}) are
dominant, which is not negligible but rather reaches several percent.
\subsection{Other helium rich white dwarfs.}

The mentioned result is important for our further research since it leads
to an expectation that the significance of the considered non-symmetric ion-atom
absorption processes could be much greater in the cases of the atmospheres of some
other helium-rich white dwarfs. We mean the atmospheres with the same or similar
$T_{eff}$ and $\log g$, but with the values of the ratio H:He which are larger by
one or even two orders of magnitude.
\begin{figure}
\begin{center}
\includegraphics[width=\columnwidth,
height=0.75\columnwidth]{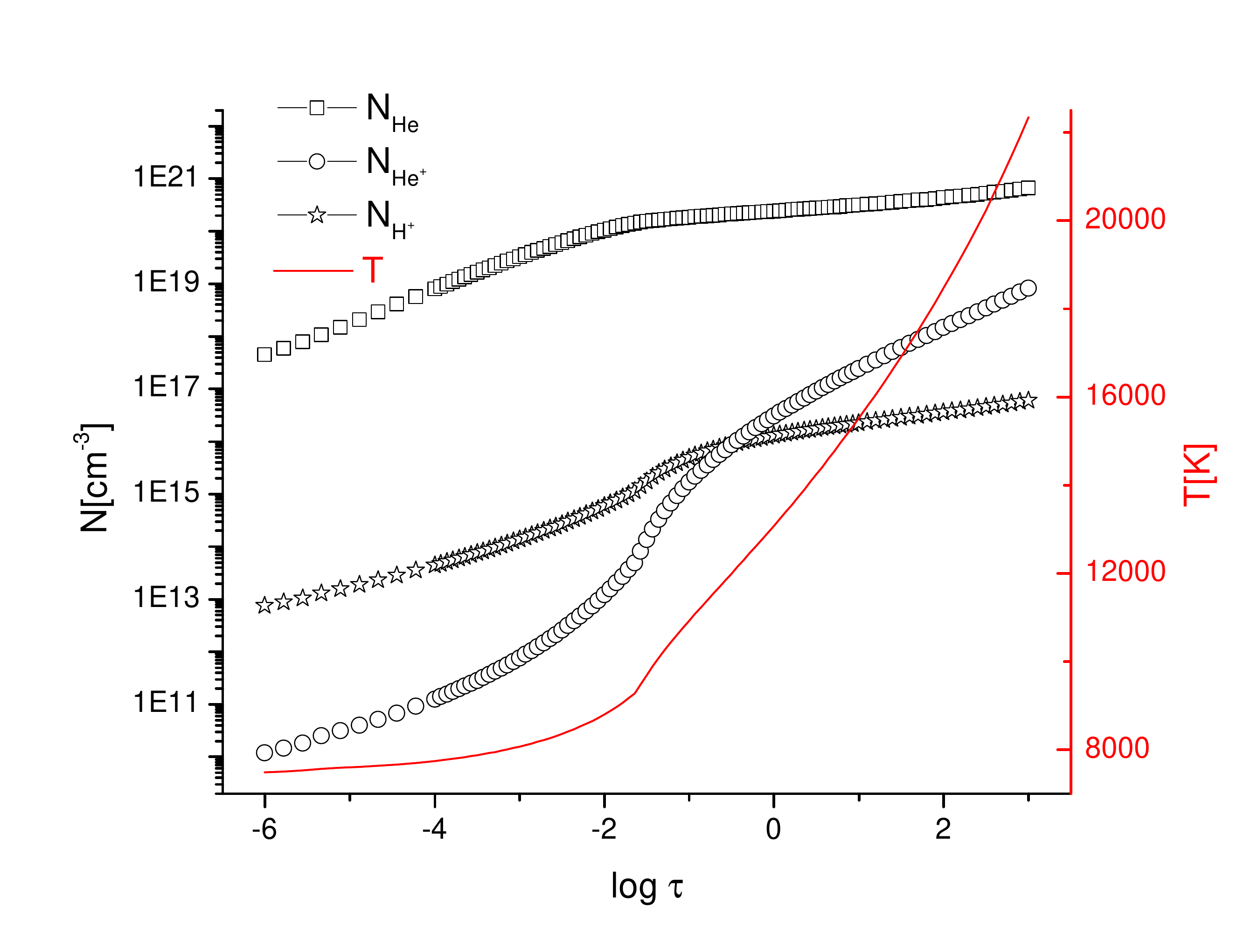} \caption{The local densities $N_{He}$,
$N_{He^+}$ and $N_{H^+}$, and the temperature $T$ as functions of $\log \tau$, where
$\tau$ is Rosseland optical depth, according to the developed model of helium-rich
white dwarf atmosphere for: $T_{eff}$=12000 K, log $g=8$ and H:He=$10^{-4}$.}
\label{fig:abund-4}
\end{center}
\end{figure}

In this context let us note that in \cite{weg85} some DC white dwarfs with
$T_{eff}\approx 12500$ K, $\log g =8$ and H:He = $2 \cdot 10^{-4}$ are described.
Then, in \cite{duf06} some weakly magnetic DZ white dwarfs with $\log g = 8$,
$T_{eff} \approx 7000$ K and H:He $\approx 10^{-3}$ are discussed. Finally, we will
remind also that in \cite{duf07} some DZ white dwarfs with $\log g \approx 8$,
$T_{eff} > 12000$ K and H:He = $10^{-4}$ and $10^{-3}$ are mentioned, as well as a
number of other DZ white dwarfs with $\log g \approx 8$, the values of $T_{eff}$ from
about $6500$ K to about $10000$ K, and the values of H:He from about $10^{-3.2}$ to
about $10^{-4.4}$. Just from the above mentioned result it follows that the
contribution of the non-symmetric processes (\ref{eq:nonsim1}) - (\ref{eq:nonsim3})
to the opacity of such atmospheres should be very significant. Namely, although the
ion H$^{+}$ density cannot increase proportionally to the ratio H:He, an
increase of this ratio of 10 or 100 times, has to cause an increase of $N_{H^{+}}$
of at least several times. So, the mentioned increase of several percent in
region $-1 < \log \tau \le 2$ in the case of DB white dwarf atmospheres has to
become an increase of several tens of percent in the cases of the helium-rich
white dwarfs with $10^{-4} \lesssim$ H:He $\lesssim 10^{-3}$.
\begin{figure}
\begin{center}
\includegraphics[width=\columnwidth,
height=0.75\columnwidth]{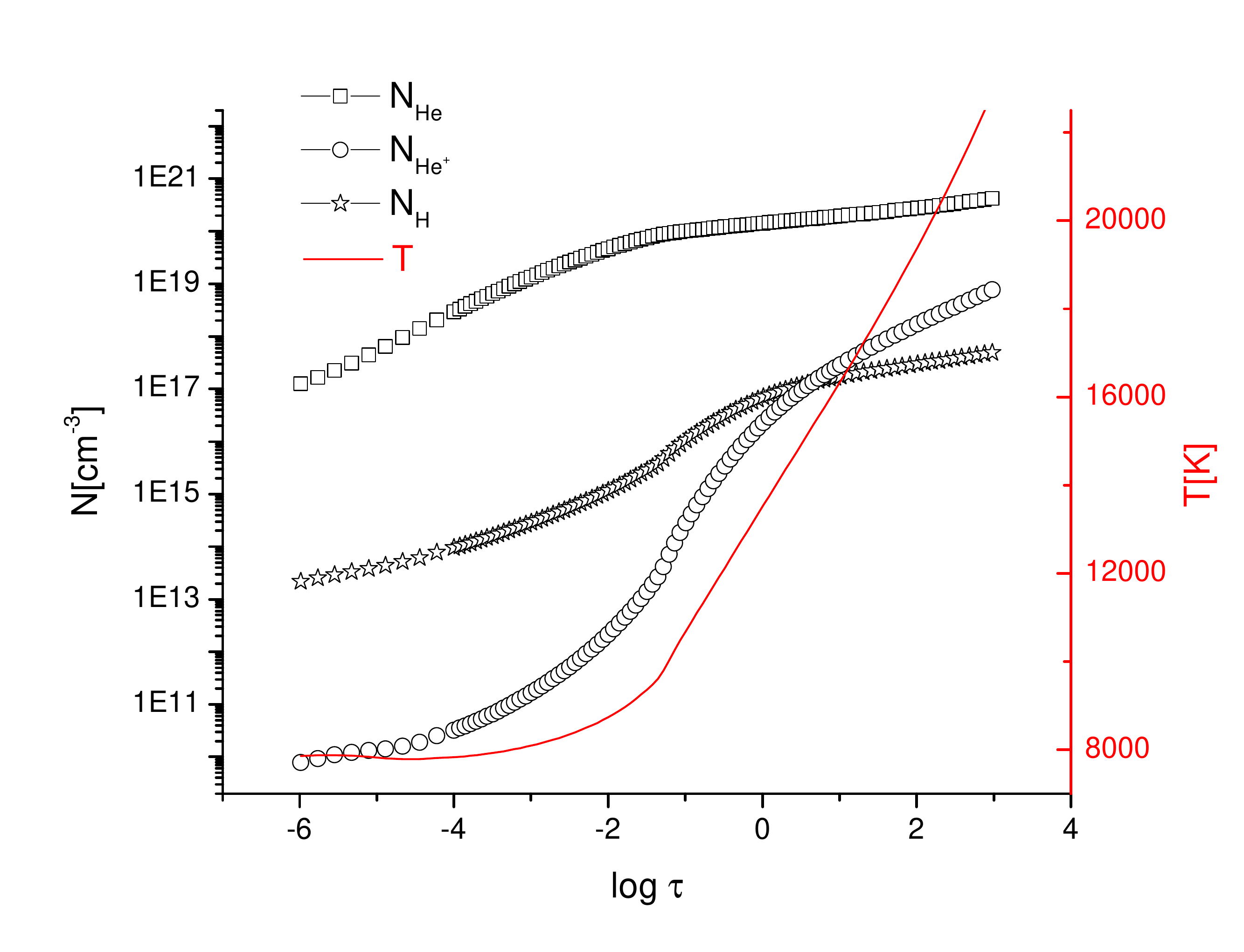} \caption{Same as in Fig. \ref{fig:abund-4},
but for H:He=$10^{-3}$.}
\label{fig:abund-3}
\end{center}
\end{figure}

In order to check our expectations we performed calculations of the quantity
$G^{(nsim)}_{ia}$, as well as of the quantities $(\lambda;\log \tau)$, $F^{(sim)}_{e-He}(\lambda;\log \tau)$
and $F^{(ia)}_{e-He}(\lambda;\log \tau)$, simulating the behavior of $T$, $N_{He}$,
$N_{H^{+}}$ and other particle densities, in helium-rich white dwarf
atmospheres with $T_{eff}=12000$ K, log $g = 8$ and H:He $>10^{-5}$. All the needed
calculations have been performed on the basis of the models taken from \cite{koe13}.
In Figs. \ref{fig:abund-4} and \ref{fig:abund-3} the corresponding densities $N_{H^{+}}$,
$N_{He^{+}}$ and $N_{e}$ are shown as functions of $\log \tau$ for H:He $=10^{-4}$
and $10^{-3}$ respectively. One can see, by comparing these figures and Fig.
\ref{fig:abundDB}, that the considered increase of the ratio H:He indeed causes a
very significant increase of $N_{H^{+}}$.
\begin{figure}
\begin{center}
\includegraphics[width=\columnwidth,
height=0.75\columnwidth]{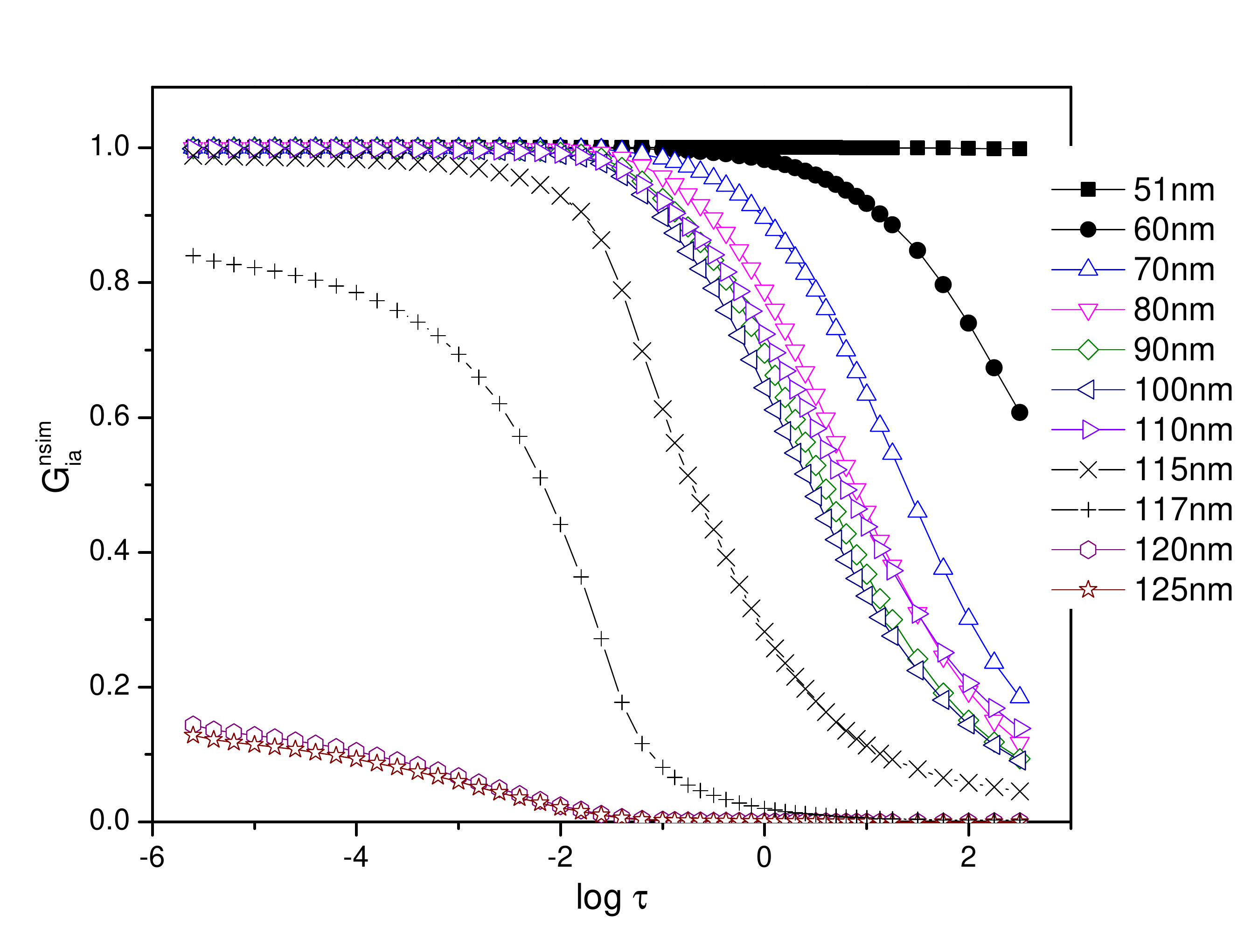} \caption{The behavior of the quantity $G_{ia}^{nsim}=\kappa_{nsim}/\kappa_{ia}$, see Eq. (\ref{eq:Gnsim}),
for the atmosphere of a helium-rich white dwarf with $T_{eff}$=12000 K, $\log g=8$ and H/He=$10^{-4}$.}
\label{fig:G4}
\end{center}
\end{figure}

\begin{figure}
\begin{center}
\includegraphics[width=\columnwidth,
height=0.75\columnwidth]{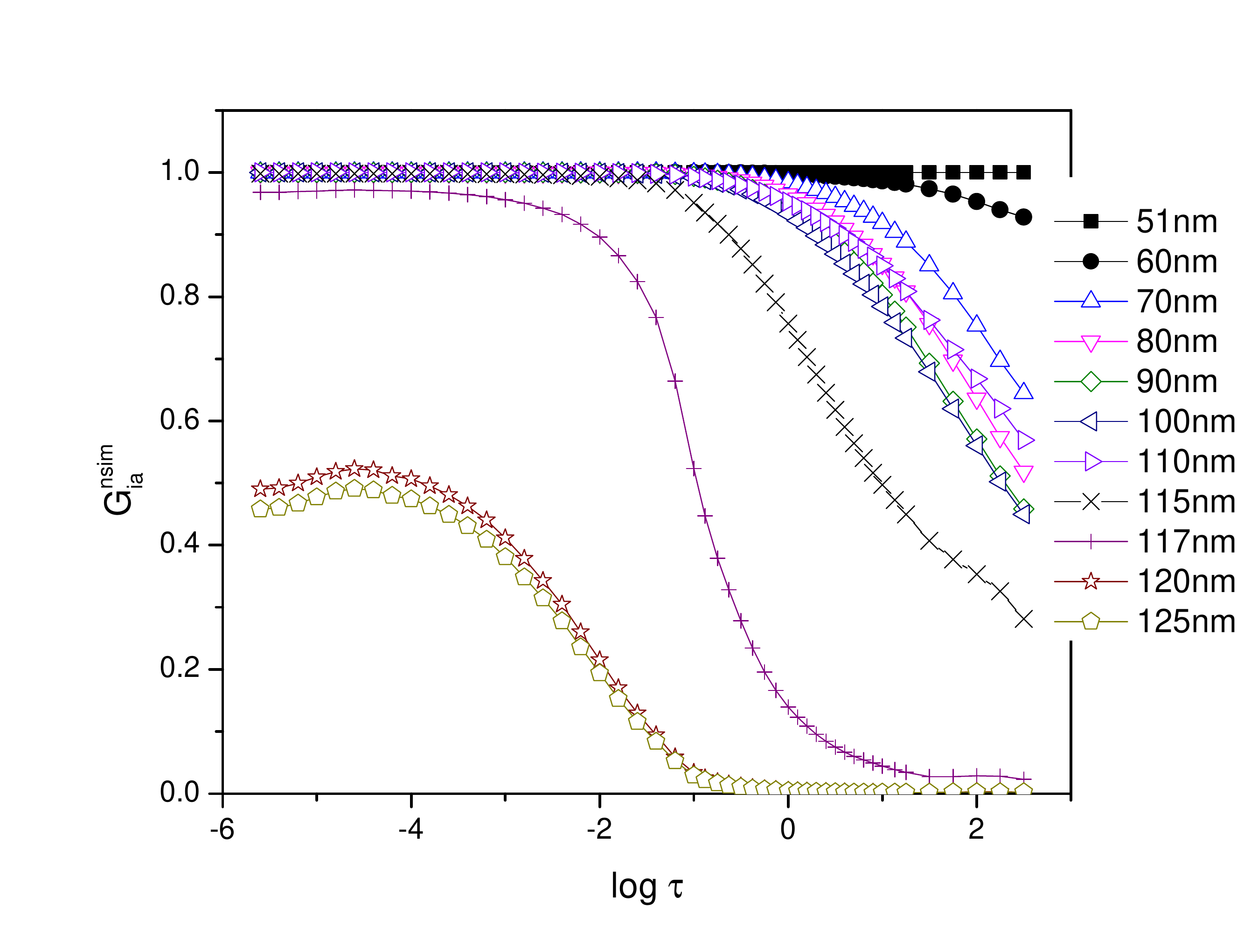} \caption{Same as in Fig. \ref{fig:G4}, but
for H:He=$10^{-3}$.}
\label{fig:G3}
\end{center}
\end{figure}

The behavior of $G^{(nsim)}_{ia}(\lambda;\log \tau)$ in the considered spectral
region (denoted by $"II"$ in Fig. \ref{fig:plank}) is shown in Fig. \ref{fig:G4} for
H:He= $10^{-4}$ and in Fig. \ref{fig:G3} for H:He=$10^{-3}$. By comparing
Fig.\ref{fig:G4} and Fig.\ref{fig:G5} it can be seen that for 51 nm $\le \lambda \le$
125 nm an increase of the ratio H:He from $10^{-5}$ to $10^{-4}$ causes a visible
increase of participation of the considered non-symmetric processes (with respect to the
total ion-atom spectral absorption coefficient) for $\log \tau >-1 $ and a significant
increase for $\log \tau >0$. Than, by comparing Fig. \ref{fig:G5} and Fig. \ref{fig:G3} it
is seen that an increase of the ratio H:He from $10^{-5}$ to $10^{-3}$ causes
yet very large increase of the said participation in the whole region
$\log \tau > -1$ .

\begin{figure}
\begin{center}
\includegraphics[width=\columnwidth,
height=0.75\columnwidth]{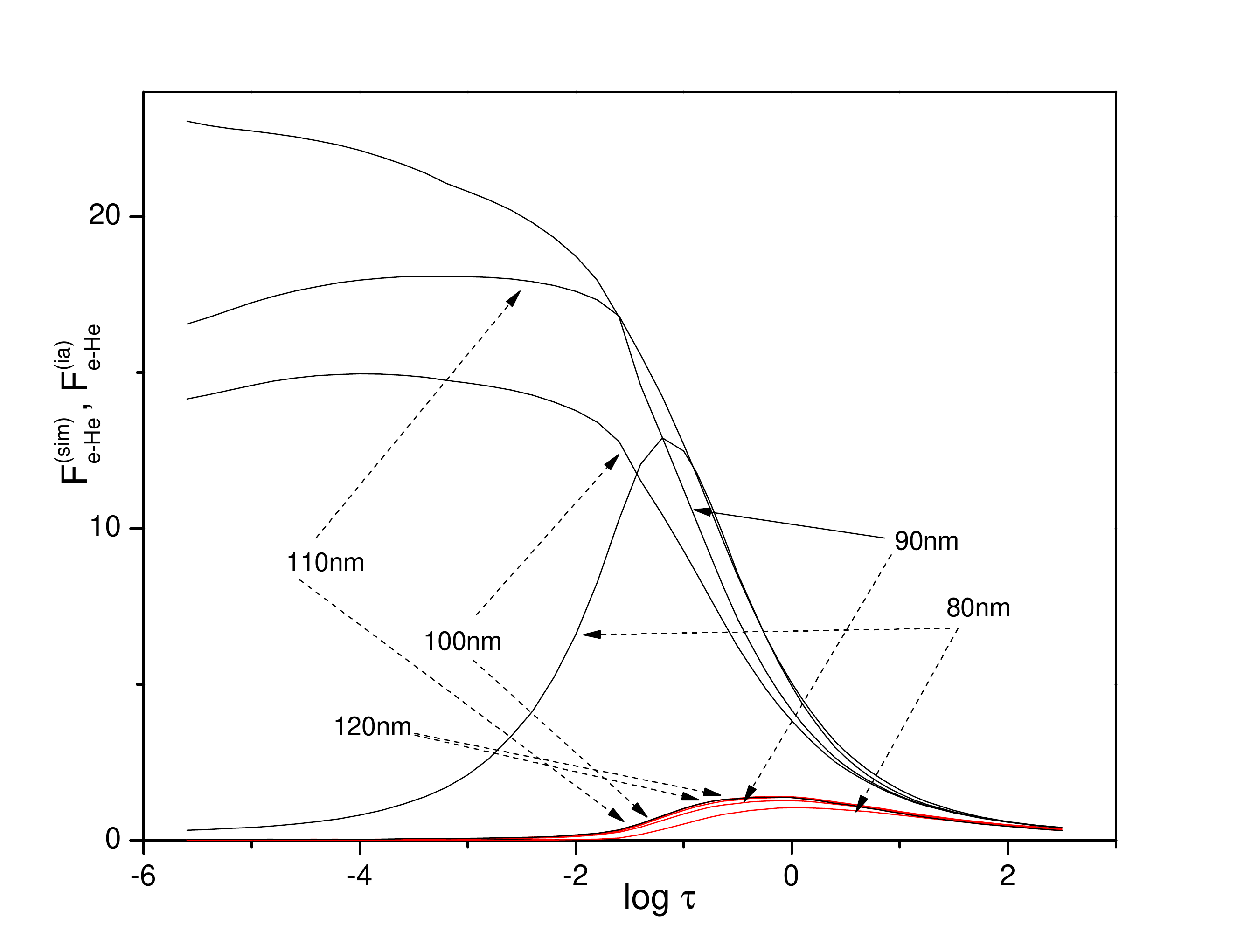} \caption{The behavior of the quantities $F^{(sim)}_{e-He}(\lambda;\log \tau)=\kappa_{sim}/\kappa_{ea}$
(red line i.e. lower line for the same $\lambda$) and $F^{(ia)}_{e-He}(\lambda;\log \tau)=\kappa_{ia}/\kappa_{ea}$(black line i.e. upper line for the same $\lambda$), see Eq. (\ref{eq:Fnsim}),
for the atmosphere of a helium rich white dwarf with $T_{eff}$=12000 K, $\log g=8$ and H/He=$10^{-4}$.}
\label{fig:F4}
\end{center}
\end{figure}
\begin{figure}
\begin{center}
\includegraphics[width=\columnwidth,
height=0.75\columnwidth]{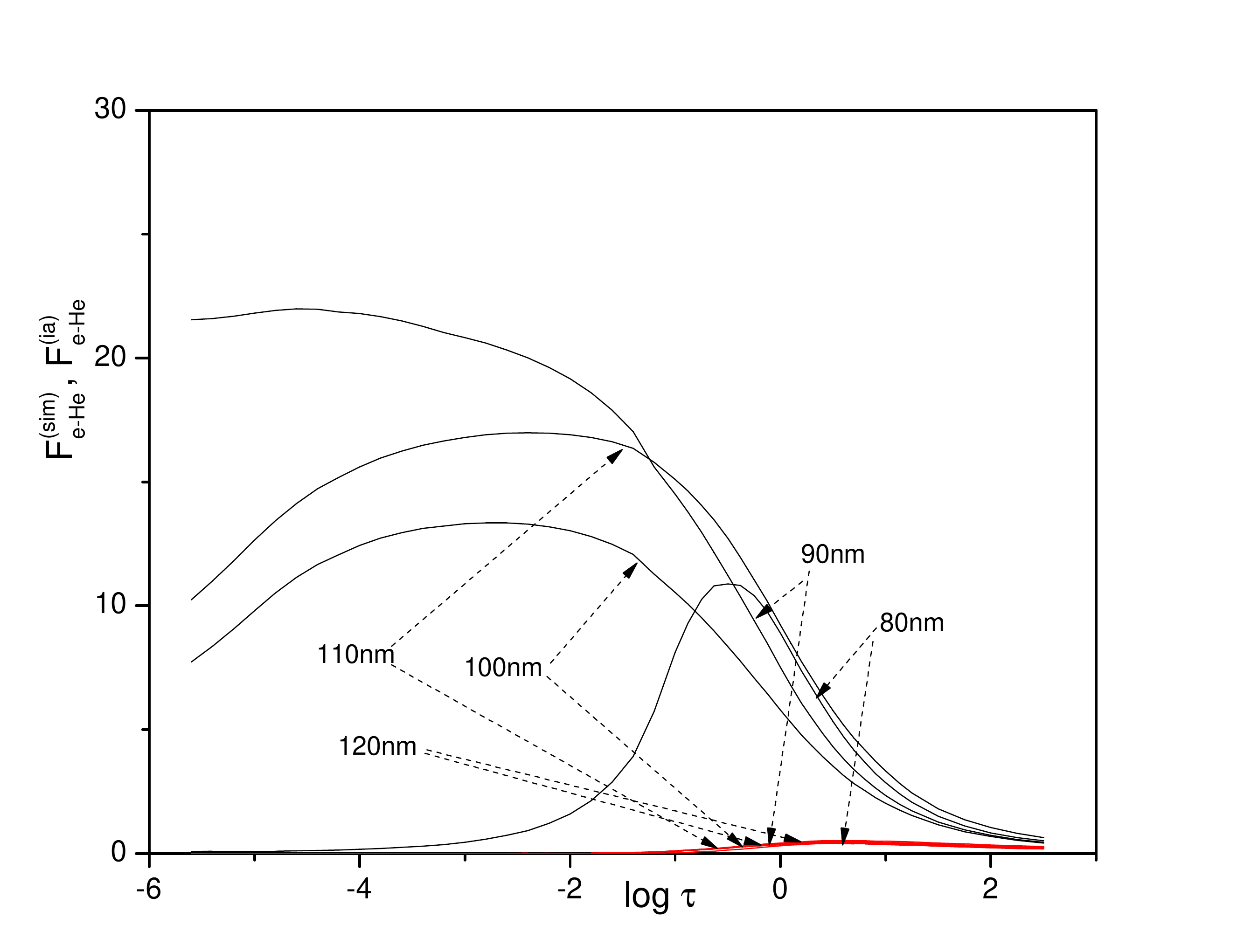} \caption{Same as in Fig. \ref{fig:F4}, but
for H:He=$10^{-3}$.}
\label{fig:F3}
\end{center}
\end{figure}

How an increase of the ratio H:He influences an increase of the relative significance
of ion-atom absorption processes with respect to the concurrent electron-atom process
(\ref{eq:e-He}) in the mentioned spectral region is illustrated by Figs.
\ref{fig:F4} and \ref{fig:F3} which show the behavior of the quantities
$F^{(sim)}_{e-He}(\lambda;\log \tau)$ and $F^{(ia)}_{e-He}(\lambda;\log \tau)$ for
H:He= $10^{-4}$ and $10^{-3}$ respectively. From these figures one can see that for
H:He= $\gtrsim 10^{-4}$ the inclusion of the non-symmetric processes
(\ref{eq:nonsim1}) - (\ref{eq:nonsim3}) causes the ion-atom absorption
processes to become absolutely dominant with respect to the electron-atom process
(\ref{eq:e-He}) in the greatest part of this region, namely for
51 nm $\le \lambda \lesssim$ 110 nm, while for $\lambda >$ 110 nm the efficiency
of ion-atom processes stays close to the efficiency of the process (\ref{eq:e-He}).

\begin{figure}
\begin{center}
\includegraphics[width=\columnwidth,
height=0.75\columnwidth]{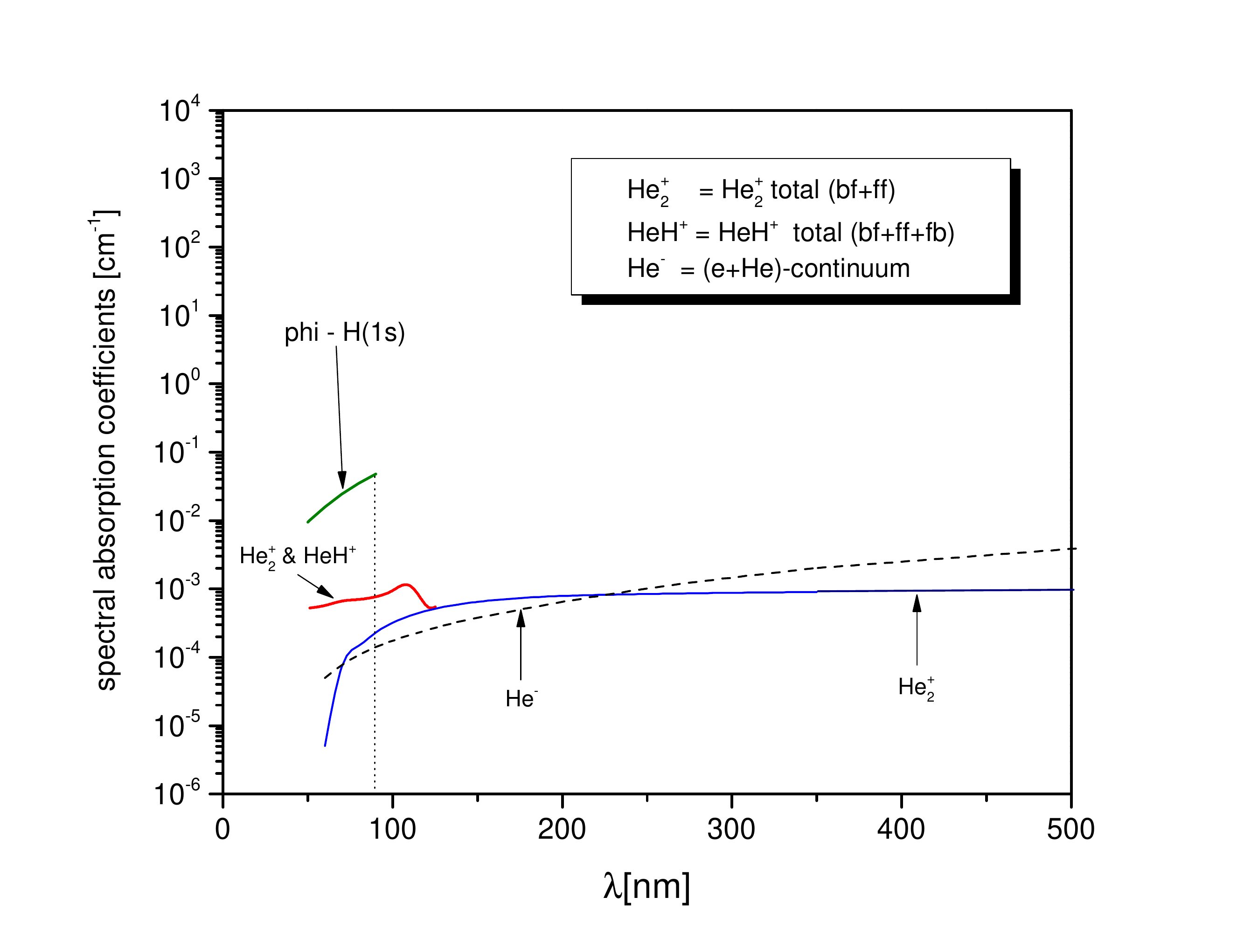} \caption{The plots of all considered absorbtion
processes for $\log \tau=0$ in the case of a helium-rich white dwarf with
$T_{eff}$=12000 K, $\log g=8$ and H:He=$10^{-4}$.}
\label{fig:2}
\end{center}
\end{figure}

\begin{figure}
\begin{center}
\includegraphics[width=\columnwidth,
height=0.75\columnwidth]{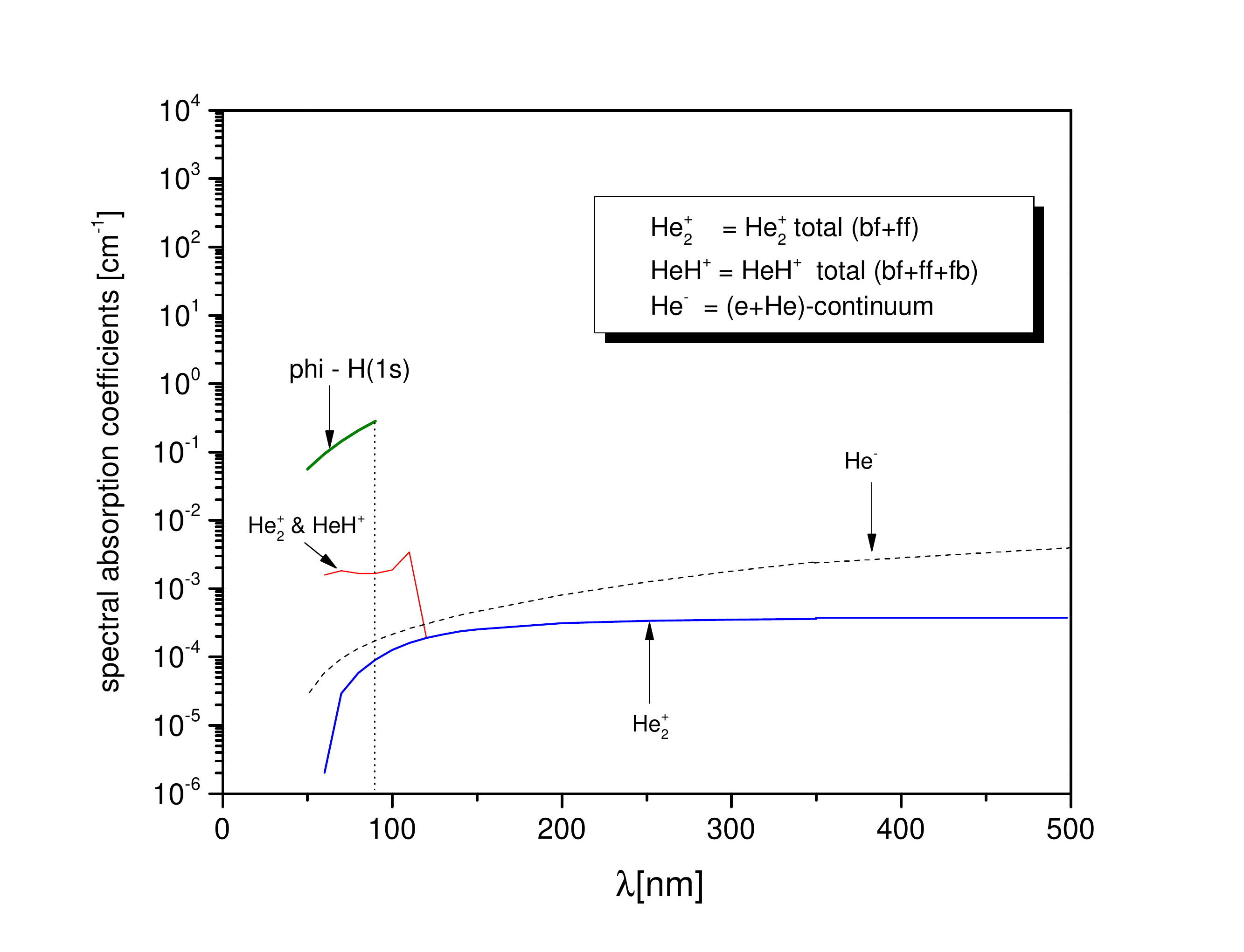} \caption{Same as in Fig. \ref{fig:2}, but
for H:He=$10^{-3}$.}
\label{fig:3}
\end{center}
\end{figure}

In order to obtain the complete picture of the discussed absorption processes in
the helium-rich white dwarf atmospheres in the cases H:He= $10^{-4}$ and $10^{-3}$
it is necessary again to include into the consideration the hydrogen photo-ionization
process \ref{eq:phi}. The significance of partial
absorption processes in such atmospheres within the whole region $\lambda > \lambda_{He}$
is illustrated in Fig's \ref{fig:2} and \ref{fig:3}  where the corresponding plots
of these processes are presented for $\log \tau=0$. One can see that, as in the case
of DB white dwarf, the process (\ref{eq:phi}) in the region
$\lambda_{He} < \lambda < \lambda_{H}$ gives the dominant contribution
to the opacity of the considered atmospheres. Then, it has been established that in
these cases this dominance also holds for any $\log \tau < 0$.

\begin{figure}
\begin{center}
\includegraphics[width=\columnwidth,
height=0.75\columnwidth]{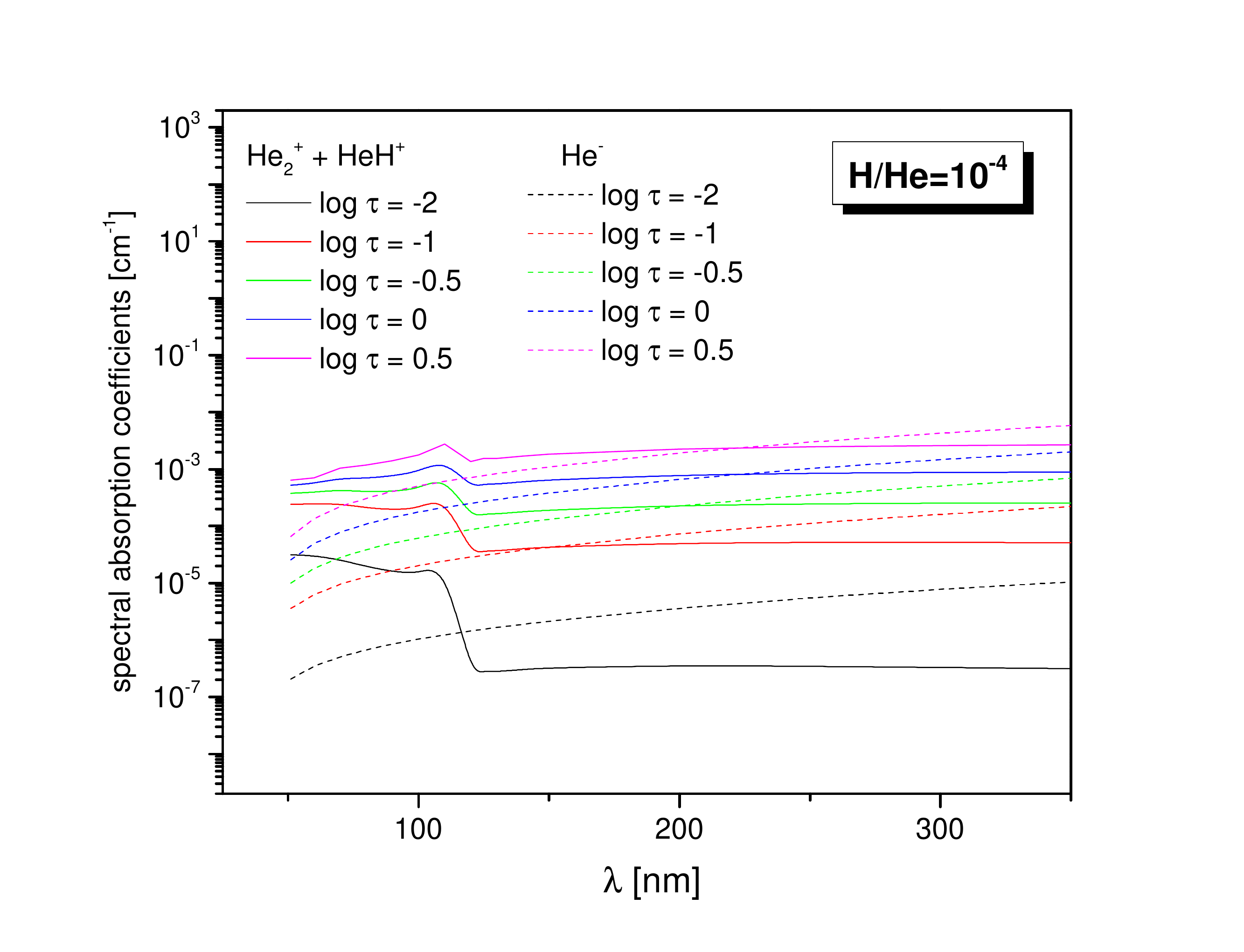} \caption{The plots of the examined ion-atom
absorbtion processes (symmetric and non-symmetric together) and the referent
electron-atom process (He$^{-}$-continuum) for $\log \tau=$ -2, -1, -0.5, 0 and 0.5
in the case of a helium-rich white dwarf with $T_{eff}$=12000 K, $\log g=8$ and
H:He=$10^{-4}$.}
\label{fig:X}
\end{center}
\end{figure}

In accordance with our considerations, it is necessary to remind
that the considered ion-atom absorption processes (symmetric and non-symmetric)
can naturally be significant in the helium-rich white dwarf atmospheres with
$T_{eff} < 20000$K, since at the higher temperatures electron-ion absorption
processes completely dominate with respect to the considered ion-atom and
electron-atom processes. Therefore, it is necessary here to take into
account the papers \cite{ber11} and \cite{vos07}, where the data about
numerous helium-rich white dwarfs are presented. Namely, from these one
can see that the values H:He which correspond to the helium-rich white
dwarfs with $T_{eff} < 20000$K are mainly situated between $10^{-5}$ and $10^{-4}$,
while the values H:He $>10^{-4}$, especially H:He close to $10^{-3}$,
have to be treated as certain extremes. But our results are supported by the following facts:\\
-firstly, from the results obtained it follows that the non-symmetric ion-atom
processes cannot be neglected already for the case of atmospheres of DB white
dwarfs with H:He=$10^{-5}$;\\
-secondly, from these results it follows that the effect of inclusion of the non-symmetric
processes in the consideration fully manifests already for
H:He = $10^{-4}$, i.e. far from the extremum (H:He=$10^{-3}$, and
remains practically the same further on. This makes important not just a neighborhood of
the extremum value H:He=$10^{-3}$, but rather the whole region of H:He$>10^{-5}$.
Since from the presented results it follows that the ion-atom absorption
processes should be especially significant in helium-rich white
dwarf atmospheres with H:He$=10^{-4}$, this case is additionally illustrated
by Fig. \ref{fig:X}. In this figure the plots of the examined ion-atom
absorbtion processes (symmetric and non-symmetric together) and of the referent
electron-atom process (He$^{-}$-continuum) are presented for $\log \tau=$ -2, -1, -0.5, 0 and 0.5
for the case H:He$=10^{-4}$. From Fig. \ref{fig:X} it can be clearly seen
how the contribution of the non-symmetric ion-atom processes and the total efficiency
of all the ion-atom processes with respect to the electron-atom
processes changes with a change of $\log \tau$ in the whole region $\lambda > \lambda_{He}$.

\begin{figure}
\begin{center}
\includegraphics[width=\columnwidth,
height=0.75\columnwidth]{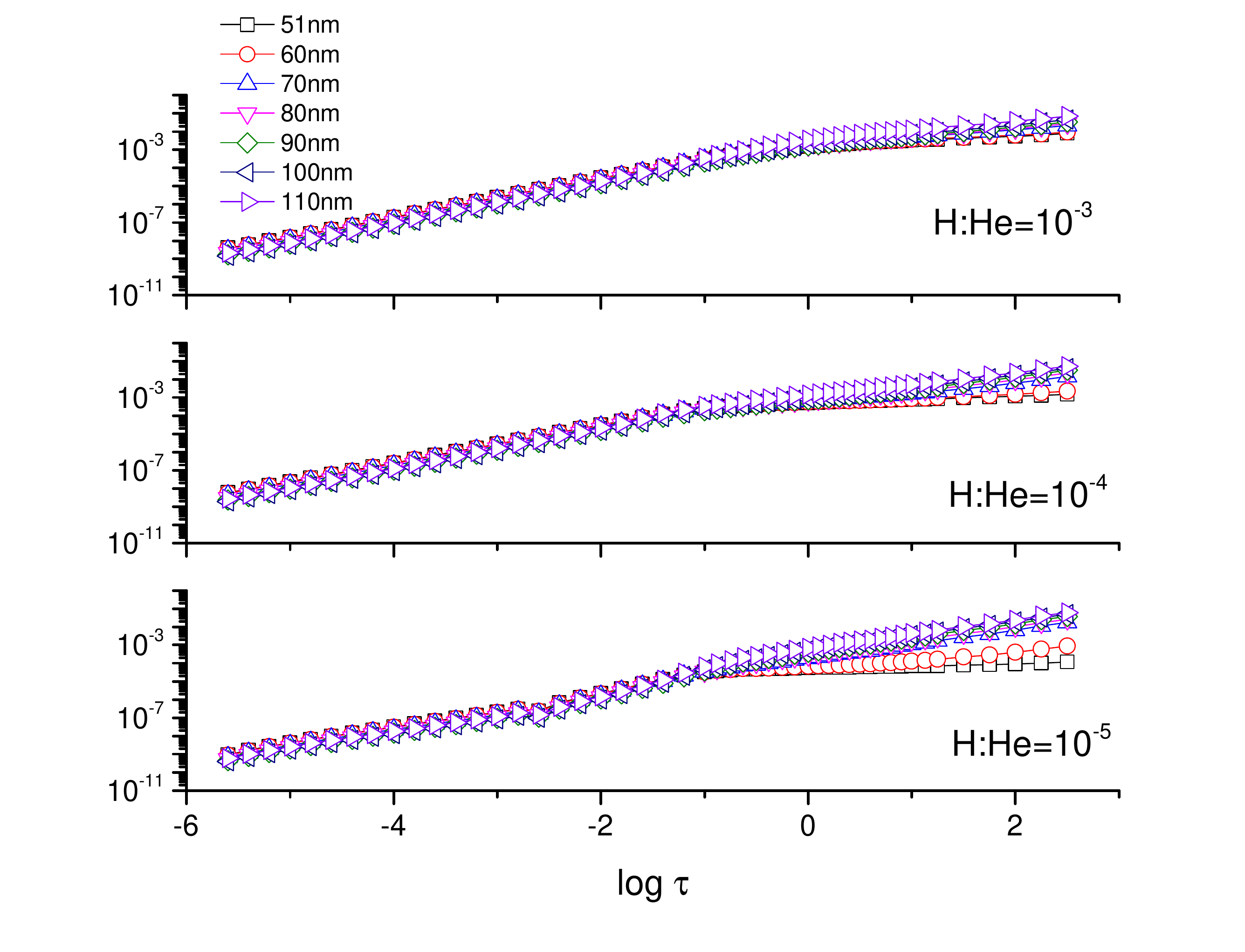} \caption{The behavior of the total
ion-atom spectral absorption coefficient $\kappa_{ia}(\lambda;\log \tau)$, see Eqs.
(\ref{eq:kapansim}), (\ref{eq:kapasim}) and (\ref{eq:kapaia}), in the cases of the
atmospheres of helium rich white dwarfs with $T_{eff} = 12000$ K and $\log g=8$
for: H:He = $10^{-5}$, H:He = $10^{-4}$ and H:He =$10^{-3}$.}
\label{fig:kappa_ia5}
\end{center}
\end{figure}

At the end of this Section in Fig. \ref{fig:kappa_ia5} the behavior is illustrated
of the total ion-atom spectral absorption coefficient
$\kappa_{ia}(\lambda;\log \tau)$, given by Eqs. (\ref{eq:kapansim}),
(\ref{eq:kapasim}) and (\ref{eq:kapaia}), for the considered examples of
atmospheres of helium-rich white dwarfs with $T_{eff} = 12000$ K and $\log g=8$:
H:He = $10^{-5}$, $10^{-4}$ and $10^{-3}$. However, for the calculations of this
absorption coefficient in the same approximation (existence of LTE), but for
different atmospheres it is necessary to know the corresponding spectral rate
coefficients, i.e. $K_{nsim}(\lambda,T)$, given by Eq. (\ref{eq:Knsim}), and
$K_{sim}(\lambda,T)$. Keeping in mind that the values of $K_{sim}(\lambda,T)$ can
be determined by means of the data from \cite{ign09}, we calculated here only the
values of the spectral rate coefficient $K_{nsim}(\lambda,T)$, within the
corresponding regions of $\lambda$ and $T$, which are presented in Table 1.


\section{Conclusions}

From the presented material it follows that the considered non-symmetric ion-atom
absorption processes have to be treated as one of the important channels of
influence on the opacity of the atmospheres of helium-rich white dwarfs in the far
UV and EUV region. So, it has been shown that even in the case of DB white dwarfs with
H:He = $10^{-5}$ such processes should be included in the models of their
atmospheres. However, the main result of this research is the establishment of the
fact that in the cases of helium-rich white dwarfs with H:He $> 10^{-5}$, and
particularly with H:He =$10^{-4}$, these non-symmetric ion-atom absorption
processes have to be included \emph{ab initio} in the models of the corresponding
atmospheres, since in the greater part of the considered far UV and EUV region they
could be completely dominant with respect to the referent electron-atom and
symmetric ion-atom absorbtion processes.

Besides, attention has been payed again in this paper to the role of the hydrogen component
in the atmospheres of helium-rich white dwarfs. Namely, it has been shown that in
all the considered cases (H:He = $10^{-5}, 10^{-4}$ and $10^{-3}$) the hydrogen
photo-ionization processes (\ref{eq:phi}) yield a dominant contribution to the opacity
of the corresponding atmospheres in the region
$\lambda_{He} < \lambda < \lambda_{H}$.

As a task for further investigations in this area the study can be mentioned of the
atmospheres of helium-rich white dwarfs with smaller effective temperatures
($\approx 7000$ K) where the significance of the hydrogen component can be greater
than in the described cases. Also, inclusion into consideration would be useful
of the ion-atom non-symmetric absorption processes with participation of some
metal components of the considered atmospheres.

\section*{Acknowledgments}

The authors are very grateful to Prof. D. Koester for providing the data of
helium-rich white dwarfs atmosphere models and to Prof. P. Bergeron for a
very wide and fruitful discussion. Also, the authors are thankful
to the Ministry of Education, Science and Technological Development of the Republic
of Serbia for the support of this work within the projects 176002 and III4402.


\newcommand{\noopsort}[1]{} \newcommand{\printfirst}[2]{#1}
  \newcommand{\singleletter}[1]{#1} \newcommand{\switchargs}[2]{#2#1}


\begin{table}
\begin{center}
\caption{The spectral absorption rate coefficient $K_{nsim}(\lambda;T)$, see
Eq. (\ref{eq:Knsim}), calculated under the condition of existence of local
thermodynamic equilibrium. }
\label{tab:rate}
\begin{tabular}
{ c c c c c c c c } \hline
    \multicolumn{7}{c}{ $T$}\\
    & & & $[10^{3}$K] & & & & \\
   \cline{2-8}
  $\lambda$ [nm]&     8&      10&      12&     14&     16&     18&    20\\
  \hline
51&3.24E-40&1.68E-40&1.04E-40&7.22E-41&5.37E-41&4.19E-41&3.39E-41\\
55&3.36E-40&1.83E-40&1.19E-40&8.58E-41&6.66E-41&5.43E-41&4.58E-41\\
60&3.19E-40&1.85E-40&1.26E-40&9.54E-41&7.70E-41&6.51E-41&5.68E-41\\
65&2.89E-40&1.78E-40&1.28E-40&1.01E-40&8.39E-41&7.29E-41&6.52E-41\\
70&2.65E-40&1.75E-40&1.33E-40&1.09E-40&9.36E-41&8.31E-41&7.54E-41\\
75&2.33E-40&1.61E-40&1.26E-40&1.06E-40&9.31E-41&8.38E-41&7.68E-41\\
80&2.09E-40&1.53E-40&1.24E-40&1.07E-40&9.61E-41&8.80E-41&8.20E-41\\
85&1.93E-40&1.48E-40&1.25E-40&1.11E-40&1.02E-40&9.48E-41&8.94E-41\\
90&1.84E-40&1.49E-40&1.30E-40&1.19E-40&1.11E-40&1.05E-40&1.00E-40\\
95&1.83E-40&1.56E-40&1.41E-40&1.32E-40&1.25E-40&1.20E-40&1.16E-40\\
100&1.91E-40&1.73E-40&1.61E-40&1.53E-40&1.47E-40&1.43E-40&1.40E-40\\
105&2.24E-40&2.16E-40&2.05E-40&1.97E-40&1.92E-40&1.90E-40&1.89E-40\\
110&3.38E-40&3.10E-40&3.07E-40&3.06E-40&3.04E-40&3.03E-40&3.02E-40\\
115&5.53E-41&6.74E-41&7.70E-41&8.65E-41&9.49E-41&1.02E-40&1.10E-40\\
120&3.14E-42&2.75E-42&2.62E-42&2.48E-42&2.25E-42&1.93E-42&1.62E-42\\
125&3.11E-42&2.58E-42&2.51E-42&2.45E-42&2.19E-42&1.75E-42&1.31E-42\\

 \hline
 \end{tabular}
 \end{center}
\end{table}

\label{lastpage}

\end{document}